\documentclass[aps,twocolumn,superscriptaddress,prl]{revtex4-1}

\usepackage{bm,graphicx,textcomp,amssymb,amsmath,dcolumn,color}

\usepackage{multirow}
\usepackage{tabularx}
\usepackage{ctable}
   
\usepackage[utf8]{inputenc}
\usepackage[T1]{fontenc}

\newcommand{\nn}{\nonumber\\}

\newcommand{\f}[1]{\mbox{\boldmath$#1$}}

\newcommand{\bea}{\begin{eqnarray}}
\newcommand{\ea}{\end{eqnarray}}
\newcommand{\eea}{\end{eqnarray}}
\newcommand{\ord}{\,{\cal O}}

\usepackage{ulem}
\usepackage{comment}
\usepackage{bbm}
\usepackage{bbold}   
\usepackage{mathtools}
\newcommand*{\I}{ {\rm i} }

\begin{document}

\title{Sauter-Schwinger effect for colliding laser pulses}

\author{Christian Kohlf\"urst}

\affiliation{Helmholtz-Zentrum Dresden-Rossendorf, 
Bautzner Landstra{\ss}e 400, 01328 Dresden, Germany,}

\author{Naser Ahmadiniaz} 
\affiliation{Helmholtz-Zentrum Dresden-Rossendorf, 
Bautzner Landstra{\ss}e 400, 01328 Dresden, Germany,} 

\author{Johannes Oertel}
\affiliation{Fakult\"at f\"ur Physik, Universit\"at Duisburg-Essen,
  Lotharstra{\ss}e 1, 47057 Duisburg, Germany,}

\author{Ralf Sch\"utzhold}

\affiliation{Helmholtz-Zentrum Dresden-Rossendorf, 
Bautzner Landstra{\ss}e 400, 01328 Dresden, Germany,}

\affiliation{Institut f\"ur Theoretische Physik, 
Technische Universit\"at Dresden, 01062 Dresden, Germany,}

\date{\today}

\begin{abstract}
Via a combination of analytical and numerical methods, we 
study electron-positron pair creation by the electromagnetic field 
$\f{A}(t,\f{r})=[f(ct-x)+f(ct+x)]\f{e}_y$ of two colliding laser pulses. 
%
Employing 
a generalized WKB approach, 
we find that the pair creation rate along the symmetry plane $x=0$ 
(where one would expect the maximum contribution) displays the same 
exponential dependence as for a purely time-dependent electric field 
$\f{A}(t)=2f(ct)\f{e}_y$. 
The pre-factor in front of this exponential does also contain corrections due 
to focusing or de-focusing effects induced by the spatially inhomogeneous 
magnetic field. 
We compare our analytical results to numerical simulations using the 
Dirac-Heisenberg-Wigner method and find good agreement. 
\end{abstract}

\maketitle

\paragraph{Introduction} 

As one of the most striking and fundamental predictions of quantum 
electrodynamics (QED), the vacuum should become unstable in the presence of 
strong electric fields, leading to the spontaneous creation of 
electron-positron pairs (``matter from light'') 
\cite{HeisenbergEuler, Weisskopf}. 
For a constant electric field $E$, the pair-creation probability $P$ 
displays an exponential dependence ($\hbar=c=1$)
\bea
\label{Schwinger}
P
\sim
\exp\left\{-\pi\,\frac{m^2}{qE}\right\}
=
\exp\left\{-\pi\,\frac{E_S}{E}\right\}
\,,
\ea
with the electron mass $m$ and elementary charge $q$, 
which can be combined to yield the Schwinger critical field 
$E_S=m^2/q\approx1.3\times10^{18}~\rm V/m$. 
The above functional dependence does not admit a Taylor expansion in $q$
which indicates that this Sauter-Schwinger effect~\eqref{Schwinger} 
is a non-perturbative phenomenon \cite{Sauter:1931, Schwinger:1951nm}.  
As a result, the corresponding calculations can be quite non-trivial and our 
knowledge beyond the case of constant fields is very limited 
\cite{PhysRevA.102.063110, PhysRevA.102.052805}. 
For slowly varying fields, we may apply the locally constant field 
approximation by evaluating Eq.~\eqref{Schwinger}, together with its 
generalization to additional
magnetic fields, at each space-time point \cite{Nikishov}.  
However, this approximation has a limited range of applicability and does not 
capture many important effects, such as the dynamically assisted 
Sauter-Schwinger effect 
\cite{DynamicallyAssistedSchwinger,Catalysis,OTTO2015335,PhysRevD.100.116018}.  

From a fundamental point of view as well as in anticipation of experimental 
initiatives aiming at ultra-high field strengths \cite{footnote1}, 
it is important to better understand the Sauter-Schwinger effect for 
non-constant fields 
\cite{PhysRevLett.101.200403, PhysRevLett.105.220407}. 
While there has been progress regarding fields which depend on one coordinate 
(e.g., space $x$ or time $t$ \cite{KimPage, Kleinert}, 
or a light-cone variable $t-x$ \cite{PhysRevD.84.125022, Heinzl}), 
our understanding of more complex field dependences, e.g., 
the interplay between spatial and temporal variations, 
is still in its infancy 
\cite{PhysRevD.72.105004, PhysRevA.99.022125}.  
Furthermore, going from simple models towards realistic field configurations
requires the consideration of transversal fields which are vacuum solutions 
of the Maxwell equations.  
In the following, we venture a step into this direction by employing a 
combination of analytical and numerical methods.  

\paragraph{The model} 

In order to treat a potentially realistic yet simple field configuration, 
we consider the head-on collision of two equal plane-wave laser pulses, 
see also \cite{aleksandrov_prd_2017_2} 
\bea
\label{collision}
\f{A}(t,\f{r})=\left[f(t-x)+f(t+x)\right]\f{e}_y
\,.
\ea
For asymmetric collision scenarios, see, e.g., 
\cite{Ilderton, Fedotov, PhysRevD.97.096004, PhysRevA.98.032121}.
This vector potential~\eqref{collision} is an even function of $x$, i.e., 
$A_y(t,x)=A_y(t,-x)$ such that $\partial_xA_y(t,x=0)=0$.
Thus, along the symmetry plane $x=0$, the electric field components $E_y$
add up while the magnetic fields $B_z$ of the two pulses cancel each other.  
As a result, one would expect the maximum contribution to pair creation there. 

In the following, we assume that the typical frequency scale $\omega$ 
describing the rate of change of the function $f(t)$ is sub-critical, 
i.e., much smaller than the electron mass $\omega\ll m$.
The characteristic electric field strength $E$ should also be sub-critical
$E\ll E_S$ and the Keldysh parameter (or inverse laser parameter $1/a_0$) 
\cite{Keldysh, Perelomov, PopovRev}
\bea 
\label{Keldysh}
\gamma=\frac{m\omega}{qE}=\frac{1}{a_0}  
\,,
\ea
should be roughly of order unity such that $qE=\ord(\omega m)$.  


\paragraph{WKB approach} 

In this limit, where the electron mass $m$ is the largest scale, we may 
employ semi-classical methods such as world-line instantons 
\cite{Affleck,PhysRevD.72.105004,Linder,PhysRevD.84.125023} discussed in 
Section~A of the Supplemental Material \cite{Supplement}
or the WKB approach \cite{PhysRevLett.113.040402,Taya} used here. 
%
%
%
For simplicity and because spin effects are not expected to play a major role 
here, we start from the Klein-Fock-Gordon equation 
\bea
\label{Klein-Fock-Gordon}
\left[
\left(\partial_\mu+iqA_\mu\right)\left(\partial^\mu+iqA^\mu\right)-m^2
\right]\phi=0
\,.
\ea
Via the standard WKB ansatz \cite{Maslov} 
\bea
\label{WBK}
\phi(t,x,y,z)=\alpha(t,x) e^{iS(t,x,y,z)}
\,,
\ea
we split $\phi$ into a slowly varying amplitude $\alpha$ and a rapidly 
oscillating phase $e^{iS}$. 
More precisely, $\partial_\mu S$ and $qA_\mu$ are large quantities of the 
order of the electron mass $\ord(m)$ while $\partial_\mu\alpha=\ord(\omega)$ 
is much smaller.
Inserting this ansatz~\eqref{WBK} into Eq.~\eqref{Klein-Fock-Gordon}, 
the leading order $\ord(m^2)$ yields the eikonal equation 
$(\partial_\mu S+qA_\mu)(\partial^\mu S+qA^\mu)=m^2$.
In view of the translational invariance in $y$ and $z$, we make the 
separation ansatz $S(t,x,y,z)=k_yy+k_zz\pm s(t,x)$,
where $s(t,x)$ is determined by the first-order equation 
\bea 
\label{square-root}
\partial_ts=\sqrt{m^2+\left(\partial_xs\right)^2+\left(k_y+qA_y\right)^2+k_z^2}
\,.
\ea
We expect the maximum contribution to pair creation along the symmetry plane 
$x=0$ where the electric field assumes its maximum, i.e., from those wave packets 
staying close to $x=0$ throughout the evolution, which implies zero momentum in 
$x$-direction $\partial_xs\big\rvert_{x=0}=0$ \cite{footnote-expect}.  
Thus (and since $A_y$ is an even function of $x$), we take $s(t,x)$ 
to be an even function of $x$ for simplicity.
%
%
After a Taylor expansion around $x=0$
\bea
\label{Taylor}
s(t,x)=s_0(t)+\frac{x^2}{2}\,s_2(t)+\ord(x^4)
\,,
\ea
we find that the zeroth order $s_0(t)$, i.e., the eikonal along $x=0$ 
is given by 
\bea 
\label{s_0}
\partial_ts_0
=
\sqrt{m^2+\left[k_y+qA_y(t,x=0)\right]^2+k_z^2}
\,,
\ea
in complete analogy to a purely time-dependent field. 

\paragraph{Focusing and de-focusing effects} 

As the next step, let us study the impact of the curvature $s_2(t)$ in 
Eq.~\eqref{Taylor}. 
Having determined the phase function $S$ by the leading-order $\ord(m^2)$
contribution to Eq.~\eqref{Klein-Fock-Gordon}, the sub-leading order 
$\ord(m\omega)$ determines the evolution of $\alpha$ via 
\bea
\label{evolution}
(\partial^\mu s)\partial_\mu\alpha=-\frac{\alpha}{2}\,\Box s
\,,
\ea
where the higher-order term $\Box\alpha=\ord(\omega^2)$ has been neglected. 
Along the symmetry plane $x=0$ where $\partial_xs=0$, 
the spatial derivative $\partial_x\alpha$ 
drops out and thus the left-hand side 
of Eq.~\eqref{evolution}
is again the same as in a purely time-dependent field. 

The right-hand side of Eq.~\eqref{evolution}, on the other hand, contains 
the additional term $\partial_x^2s\big\rvert_{x=0} =s_2$. 
This curvature contribution can be obtained by inserting Eq.~\eqref{Taylor}
into Eq.~\eqref{square-root} followed by a Taylor expansion 
\bea 
\label{curvature}
\partial_ts_2
=
\left.
\frac{s_2^2+\left[k_y+qA_y\right]q\partial_x^2A_y
}{\sqrt{m^2+\left[k_y+qA_y\right]^2+k_z^2}} 
\right\rvert_{x=0}
\,. 
\ea
In analogy to Eq.~\eqref{s_0}, we obtain a closed ordinary differential 
equation for $s_2(t)$. 
In contrast to Eq.~\eqref{s_0}, however, this is a non-linear equation 
which can display (blow-up) singularities.
Similar to caustics,
they do not imply singularities of the solutions $\phi$ to the original 
(linear) Klein-Fock-Gordon equation~\eqref{Klein-Fock-Gordon}, but indicate 
a break-down of the WKB ansatz~\eqref{WBK}, as also discussed in 
\cite{Oertel}.
Fortunately, for a large class of parameters including the cases of interest 
here, such singularities do not occur -- see also Section~F 
in the Supplemental Material \cite{Supplement}.
%
%

In order to provide an intuitive interpretation of the above 
equation~\eqref{curvature}, 
we note that $k_y+qA_y$ is the mechanical momentum in $y$-direction, 
proportional to the velocity $v_y$. 
As $\partial_xA_y$ is the magnetic field $B_z$, the numerator 
in Eq.~\eqref{curvature} yields, apart from the non-linearity $s_2^2$, 
the divergence $\partial_xF_x$ of the Lorentz force.  
Thus the curvature $s_2$ is associated to the focusing or de-focusing 
effect of the inhomogeneous magnetic field $B_z$. 

\paragraph{Particle creation} 

The simple WKB ansatz~\eqref{WBK} is not well suited for studying pair 
creation because this phenomenon is associated with a mixing of positive 
and negative frequency solutions, which is not captured by the 
ansatz~\eqref{WBK} for slowly varying $\alpha$.
Thus, we adapt a generalized WKB ansatz, see also 
\cite{Oertel,PhysRevD.83.065028, Popov}.

To this end, we define the phase-space
pseudo-vector $\f{\varphi}=(\phi,\dot\phi)^T$
which allows us to cast the original second-order 
equation~\eqref{Klein-Fock-Gordon} into a first-order form 
\bea
\label{first-order}
\partial_t\f{\varphi}
&=&
\left(
\begin{array}{cc}
0 & 1 \\ \partial_x^2-\mu^2 & 0
\end{array}
\right)\cdot\f{\varphi}
\nn
&=&
\left[\f{\sigma}_++\f{\sigma}_-\left(\partial_x^2-\mu^2\right)\right]
\cdot\f{\varphi}
\,,
\ea
where $\f{\sigma}_\pm$ are the Pauli ladder matrices and $\mu(t,x)$ denotes 
the effective mass $\mu^2=m^2+\left(k_y+qA_y\right)^2+k_z^2$. 

In order to include pair creation, we generalize the original 
WKB ansatz~\eqref{WBK} via 
\bea
\label{Bogoliubov}
\f{\varphi}=\alpha\f{u}_+e^{+is}+\beta\f{u}_-e^{-is}
\,,
\ea
where $\alpha(t,x)$ and $\beta(t,x)$ are the Bogoliubov coefficients,
which are assumed to be slowly varying. 
The basis vectors $\f{u}_\pm(t,x)$ are eigenvectors of the matrix 
\bea
\left[\f{\sigma}_+-\f{\sigma}_-\left([\partial_xs]^2+\mu^2\right)\right]
\cdot\f{u}_\pm
=
\pm i \chi \f{u}_\pm
\,,
\ea
with eigenvalues $\pm i\chi$ where $\chi(t,x)=\partial_ts(t,x)$ 
is given by Eq.~\eqref{square-root}. 
Thus, after inserting the generalized ansatz~\eqref{Bogoliubov} into 
Eq.~\eqref{first-order}, the leading order again corresponds to the 
eikonal equation~\eqref{square-root}. 

For simplicity, we use the (non-normalized) eigenvectors 
$\f{u}_\pm=(1,\pm i\chi)^T$ in the following. 
Since $A_y$ and $s$ are even functions of $x$, the first $x$-derivatives 
of $s$, $\mu$, $\chi$ and $\f{u}_\pm$ vanish along the symmetry plane $x=0$.
Furthermore, although the second $x$-derivatives of $\mu$, $\chi$, 
$\f{u}_\pm$, $\alpha$ and $\beta$ do not vanish along the symmetry plane $x=0$,
they scale with $\ord(\omega^2)$.
Thus, they are neglected within the next-to-leading order $\ord(m\omega)$ 
of the WKB approach, which yields (along the symmetry plane $x=0$)
\bea
\Big(
\dot\alpha\f{u}_++\alpha\dot{\f{u}}_+-i\alpha\f{\sigma}_-\cdot\f{u}_+
\partial_x^2s
\Big)e^{+is}
+
\nn
\Big(
\dot\beta\f{u}_-+\beta\dot{\f{u}}_-+\beta\f{\sigma}_-\cdot\f{u}_-
\partial_x^2s
\Big)e^{-is}
=0
\,.
\ea
Note that $\partial_x^2s=\ord(m\omega)$ is kept, in complete analogy 
to Eq.~\eqref{evolution}. 
Projection with $\f{u}_\pm^\perp=(\pm i\chi,1)^T$ gives 
\bea
\label{evolution-Bogoliubov}
2\chi\dot\alpha+
\alpha\Box s
&=&
\beta(\Box s)e^{-2is}
\,,
\nn
2\chi\dot\beta+
\beta
\Box s
&=&
\alpha(\Box s)e^{+2is}
\,.
\ea
For the spatially homogeneous limit where $\partial_x^2s=0$, we recover the 
well-known evolution equations for a purely time-dependent field as
$\Box s\to\ddot s$. 
For our colliding-pulse scenario~\eqref{collision}, these two evolution 
equations~\eqref{evolution-Bogoliubov} for $\alpha$ and $\beta$
along the $x=0$ plane contain 
the same exponents $e^{\pm is}$ as in the case of a purely time-dependent 
field, the only difference are the pre-factors $\Box s$ which now contain 
the additional $\partial_x^2s$ term.
%
%
The $\ddot s$ 
contribution 
$\partial_t^2s=\partial_t\chi=q\dot A_y(k_y+qA_y)/\chi$ 
already present in a purely time-dependent scenario 
contains the electric field $E_y$
while the additional $\partial_x^2s$ contribution stems from the 
inhomogeneities of the magnetic field $B_z$ and describes the 
focusing or de-focusing effects, see the discussion below. 

As in the purely time-dependent scenario, we may combine the two 
linear evolution equations~\eqref{evolution-Bogoliubov} for the 
Bogoliubov coefficients into a single Riccati equation 
$\dot R=\Box s(e^{+2is}-R^2e^{-2is})/(2\chi)$
for their ratio $R=\beta/\alpha$. 

\paragraph{Numerical simulations} 

Let us compare our analytical findings with numerical simulations.
Numerical approaches to the Sauter-Schwinger effect include direct 
integrations of the Klein-Fock-Gordon or Dirac equations 
(see, e.g., 
\cite{Ruf, aleksandrov_prd_2016, PhysRevA.97.022515, PhysRevA.81.022122}),
a reformulation in terms of the Heisenberg-Wigner formalism (see, e.g., 
\cite{Kohlf,PhysRevD.101.096009, PhysRevD.87.105006}),
quantum Monte-Carlo methods (see, e.g., \cite{Klingmuller}), 
or numerical world-line instanton solvers (see, e.g., 
\cite{Schneider,PhysRevD.98.085009}).
Each of these methods has advantages and drawbacks, but calculating an 
exponentially small pair-creation probability $P$ in a complex 
higher-dimensional field configuration $\f{A}(t,\f{r})$ is always challenging. 

In order to reduce the computational complexity as much as possible, 
we consider the Dirac equation in 2+1 dimensions, where we can use 
two-component spinors, but still incorporate a transversal 
field~\eqref{collision}. 
Employing the Dirac-Heisenberg-Wigner formalism, the problem is mapped 
onto a set of first-order transport equations involving bi-linear 
expectation values, see Section~B in the Supplemental Material \cite{Supplement}.  

We consider the following field profile in Eq.~\eqref{collision}
\bea
\label{Gaussian}
f(t)=\frac{Et}{2}\,\exp\left\{-\omega^2t^2\right\}
\,,
\ea
which displays the maximum electric field $E$ at $t=0$ and $x=0$.
Since the vector potential vanishes asymptotically $f(t\to\pm\infty)=0$ 
the wavenumber $k_y$ coincides with the 
mechanical momentum at those times. 
This simplifies the numerical analysis and
will be relevant for the pair-creation spectra discussed 
in Section~D in the Supplemental Material \cite{Supplement}. 

\paragraph{Numerical results} 

In the following, we set the field parameter $E$ in Eq.~\eqref{Gaussian} to 
$E=E_S/3$, i.e., the peak field strength is one third of the Schwinger 
critical field. 
In this case, we are already in the sub-critical regime where the 
pair-creation probability $P$ is exponentially suppressed as in 
Eq.~\eqref{Schwinger}, but the numbers are not too small for a reliable 
numerical computation. 

\begin{figure}[ht]
\includegraphics[trim={1cm 0cm 3cm 1cm},clip,width=0.49\textwidth]
{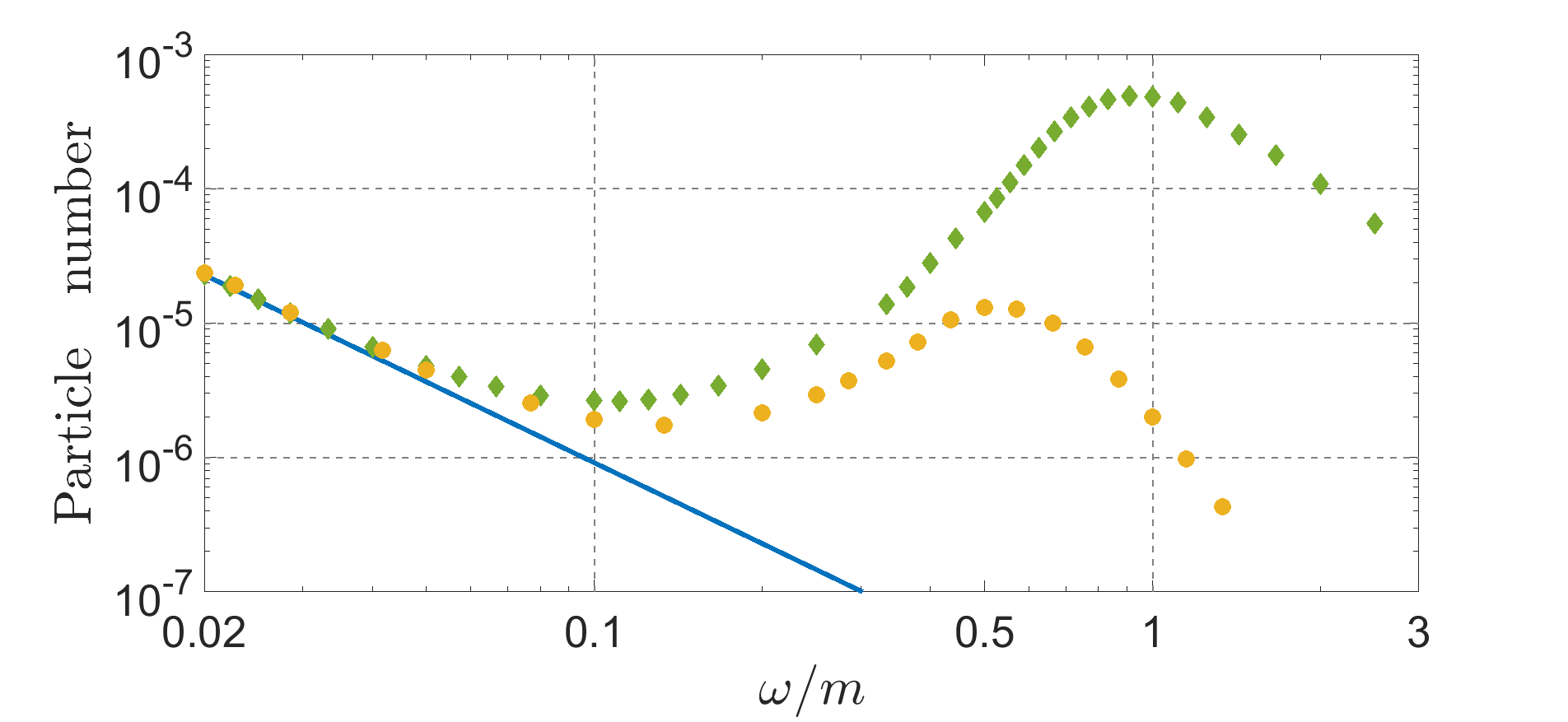} 
\caption{Plot of the mean number of created particles as a function of 
$\omega$ for the profile~\eqref{Gaussian} with $E=E_S/3$. 
The yellow circles denote the results of the Dirac-Heisenberg-Wigner 
formalism, the green diamonds correspond to the spatially homogeneous field 
approximation and the blue line displays the locally constant field 
approximation. 
}
\label{vergleich}
\end{figure}

The computed mean particle numbers are plotted in Fig.~\ref{vergleich}. 
The locally constant field approximation just reflects the trivial 
space-time volume scaling with $1/\omega^2$. 
As expected, the results of the Dirac-Heisenberg-Wigner formalism 
converge to that approximation for small $\omega$, i.e., 
small Keldysh parameters~\eqref{Keldysh}, but start to show 
significant deviations for Keldysh parameters of order unity, 
which is the regime we are interested in.

Motivated by the above findings based on 
the WKB approach, we also compared those results with the spatially 
homogeneous field approximation: 
To this end, we calculated the pair-creation probability $P$ for a purely 
time-dependent scenario $\f{A}(t)=2f(ct)\f{e}_y$ corresponding to the 
field at the symmetry plane $x=0$ 
\cite{Hebenstreit, Kluger, Schmidt}. 
While this is expected to yield the correct pair-creation exponent, 
this scenario grossly overestimates the pre-factor because particles are 
now created in the whole spatial volume. 
For the colliding pulses~\eqref{collision}, 
however, pair creation predominantly occurs in 
the vicinity of the symmetry plane $x=0$ where the electric field 
assumes its maximum. 
In order to correct this over-estimation, we introduce a pre-factor accounting
for the finite extent (in $x$-direction) of the effective pair-creation volume 
\cite{footnote2}. 
As a natural and minimal assumption, we take this pre-factor to be proportional 
to $1/\omega$, i.e., the pulse width, where the proportionality constant is 
fixed by demanding convergence to the locally constant field approximation at 
small $\omega$, see also 
\cite{PhysRevD.93.025014, PhysRevD.91.045016, STROBEL20141153,PhysRevD.99.016020,PhysRevD.78.061701}.

As we may observe in Fig.~\ref{vergleich}, this spatially homogeneous field 
approximation still over-estimates the pair-creation probability a bit, but 
provides a much better description than the locally constant field 
approximation.
Even for frequencies of the order of the electron mass, it reproduces the 
qualitative behavior of the full Dirac-Heisenberg-Wigner results, such as 
the peak of the particle number at $\omega=\ord(m)$. 
The quantitative disagreement regarding the height and location of the peaks 
can presumably be explained by a threshold effect marking the transition from 
the non-perturbative to the perturbative regime at large $\omega$ 
(where the 
WKB approach is expected to break down), 
see Section~E in the Supplemental Material \cite{Supplement}.

\paragraph{Focusing/de-focusing corrections} 

The spatially homogeneous field approximation explained above does only take 
into account the $\ddot s$ term in Eqs.~\eqref{evolution-Bogoliubov}, i.e., 
the electric field $E_y$. 
In order to include the effects of the magnetic field $B_z$, 
one should replace $\ddot s\to\Box s$, cf.~Eqs.~\eqref{evolution-Bogoliubov}, 
which also contains 
$\partial_x^2s$, i.e., the curvature $s_2$ in Eq.~\eqref{curvature}. 
The effect of this replacement can be studied by numerically solving the 
set of ordinary differential equations~\eqref{s_0}, \eqref{curvature} and 
\eqref{evolution-Bogoliubov} for the profile~\eqref{Gaussian}. 
As example parameters, we choose $E=E_S/3$ as before and $\omega=m/3$,
i.e., $\gamma=1$. 

As shown in Section~C of the Supplemental Material \cite{Supplement}, the 
behavior of $\ddot s_0(t)$ and $s_2(t)$ strongly depends on the momentum $k_y$.
For $k_y=\pm m$, for example, the curvature $s_2(t)$ is quite close to 
$\ddot s_0(t)$ thus almost canceling each other in the pre-factor $\Box s$.
For $k_y=0$, this is not the case as the curvature $s_2(t)$ varies more 
slowly with time than $\ddot s_0(t)$. 

We find that including the curvature term $s_2$ reduces the pair-creation 
probability, e.g., roughly by a factor of two for the case $k_y=0$ 
(which yields the dominant contribution),
see the Supplemental Material \cite{Supplement}.
Thus, including the focusing/de-focusing effects corrects the 
over-estimate of the spatially homogeneous field approximation 
and brings the estimated pair-creation probability almost on top of 
the value obtained by the Dirac-Heisenberg-Wigner approach.
However, more systematic investigations are needed to assess the 
overall accuracy of this approach.

\paragraph{Conclusions} 

As a prototypical example for a space-time dependent and transversal field 
configuration (as a vacuum solution to the Maxwell equations), we consider 
the head-on collision of two plane-wave laser pulses. 
Via the 
WKB approach, we study 
electron-positron pair creation in this background for sub-critical fields 
$E\ll E_S$ and Keldysh parameters of order unity. 
Along the symmetry (i.e., collision) plane, where we expect that dominant
contribution, we find that the pair-creation exponent is the same 
as for a purely time-dependent electric field, only the pre-factor 
$\ddot s\to\Box s$ does also include the impact of the magnetic field,  
leading to focusing/de-focusing effects. 

This approximate mapping to a purely time-dependent electric field allows us 
to employ the spatially homogeneous field approximation, which we compare to 
numerical simulations using the Dirac-Heisenberg-Wigner approach. 
We find that the spatially homogeneous field approximation over-estimates 
the pair-creation probability slightly, but provides a much better description
than the locally constant field approximation, see Fig.~\ref{vergleich}. 
It even reproduces qualitative features of the pair-creation spectra,  
see Section~D in the Supplemental Material \cite{Supplement}.

Going beyond the spatially homogeneous field approximation, we may also study 
the impact of the magnetic field, leading to focusing/de-focusing effects. 
Along the symmetry plane, this amounts to replacing $\ddot s$ by $\Box s$ 
in the evolution equations for the Bogoliubov coefficients, which also 
contains the curvature term $\partial_x^2s$. 
For the cases we studied, we found that this replacement tends to lower the 
pair-creation probability, which brings it closer to the results of the 
Dirac-Heisenberg-Wigner approach. 

However, one might also imagine other scenarios. 
Note that $\ddot s$ is
a local function of $A_y$ and $\dot A_y$, 
while the curvature $\partial_x^2s$ 
is non-local, i.e., depends on whole history of the evolution. 
This could be exploited in pulse-shape optimization schemes aimed at 
increasing the pair-creation probability. 
As an intuitive picture, if the initial wave-packet of the fermionic quantum 
vacuum fluctuations is focused onto the symmetry plane, where the electric 
field assumes its maximum, it can react to this strong field 
(i.e., produce particles) much better than a wave-packet which is more de-localized. 

\paragraph{Experimental scenarios}

Finally, let us discuss potential experimental tests of our results. 
Ultra-strong optical laser foci have very small Keldysh parameters 
$\gamma\ll1$ and should thus be treatable via the locally constant 
field approximation.
X-ray free electron lasers (XFEL), on the other hand, have much larger 
$\gamma$ and could require going beyond that approximation.
Unfortunately, however, present-day facilities do not reach the necessary 
field strengths $E$ yet. 
An interesting idea to achieve this goal is high-harmonic focusing
(see, e.g., 
\cite{PhysRevLett.94.103903, PhysRevLett.123.105001, PhysRevLett.127.114801}) 
which typically also corresponds to non-negligible $\gamma$. 

As a completely different scenario for generating ultra-strong fields, 
collisions of heavy nuclei have been studied theoretically and experimentally, 
see, e.g., 
\cite{PhysRevLett.127.052302, Reinhardt, Schweppe, Cowan, Ahmad, Maltsev}.
Considering ultra-peripheral ``collisions'' at relativistic velocities $v$ 
along the trajectories $\f{r}(t)=\pm(vt,b/2,0)^T$ with impact parameters $b$, 
the superpositions of the boosted Coulomb fields of the two nuclei 
can be approximated by Eq.~\eqref{collision} at sufficiently large distances 
$|\f{r}|\gg b$, say, of the order of the Compton length 
\cite{electron-bunch, magnetic-monopole}. 
The associated field strengths may reach or even exceed the Schwinger 
critical field $E_S$ \cite{HeavyIon,ATLAS} 
and the Keldysh parameters $\gamma$ will also be 
non-negligible (especially for ultra-relativistic $v$).  
Of course, the field strengths and their spatial and temporal gradients will 
be even larger at smaller distances $|\f{r}| \sim b$, 
such that the total electron-positron 
yield will also contain contributions from this region. 
Nevertheless, this again shows the importance of understanding the impact of 
space-time dependence on pair creation, i.e., 
to go beyond the locally constant field approximation.

\acknowledgments 

\paragraph{Acknowledgments}
 
We thank Christian Schneider and Christian Schubert for fruitful discussions. 
R.S.~acknowledges support by the Deutsche Forschungsgemeinschaft 
(DFG, German Research Foundation) -- Project-ID 278162697 -- SFB 1242.


\clearpage
\onecolumngrid
\appendix
\setcounter{equation}{0}
\setcounter{figure}{0}
\setcounter{table}{0}
\renewcommand{\theequation}{S\arabic{equation}}
\renewcommand{\thefigure}{S\arabic{figure}} 
\renewcommand{\bibnumfmt}[1]{[S#1]}
\renewcommand{\citenumfont}[1]{S#1} 

\section{A. World-line instanton technique} 

It might be illuminating to study the colliding-pulse scenario in Eq.~(2) 
via the world-line instanton technique. 
This method is another semi-classical approach and allows us to 
%
%
estimate the pair-creation probability $P$ via \cite{Affleck, PhysRevD.72.105004}
\bea
\label{instanton}
P\sim\exp\left\{-{\cal A}_{\rm inst}\right\} 
\,,
\ea
where ${\cal A}_{\rm inst}$ is the action of the associated 
world-line instanton, i.e., a closed loop $x^\mu(\tau)$
in Euclidean space-time as a solution of the Euclidean 
semi-classical or classical equations of motion 
\bea 
\label{Euclidean}
m\frac{d^2x^\mu}{d\tau^2}=qF^{\mu\nu}\frac{dx_\nu}{d\tau}
\,,
\ea
parametrized by the ``proper'' time $\tau$. 

As explained after Eq.~(2), 
the magnetic field vanishes in the $x=0$ plane and 
thus the Euclidean field strength tensor 
$F^{\mu\nu}(x=0)$ does only contain the electric field $E_y$. 
As a result, the instanton moves in imaginary time and $y$-direction, 
but stays on the $x=0$ plane, which yields the same exponent 
${\cal A}_{\rm inst}$ as for a purely time-dependent field 
$\f{A}(t)=2f(ct)\f{e}_y$, see also 
\cite{Linder,PhysRevD.84.125023}. 

Within the world-line instanton technique, the pre-factor in front of the 
exponential~\eqref{instanton} can be obtained (at least in principle) by 
considering perturbations around the instanton trajectory, but such a 
calculation can be quite non-trivial for genuinely space-time-dependent fields 
\cite{PhysRevD.73.065028,footnote-prefactor}. 

\section{B. Dirac-Heisenberg-Wigner approach}
 
Let us provide a brief outline of the Dirac-Heisenberg-Wigner approach 
\cite{Vasak:1987umA}.
We start from the Lagrangian
\begin{equation}
{\mathcal L} = 
\frac{1}{2} \left( 
\I \bar{\Psi} \gamma^{\mu} \mathcal{D}_{\mu} \Psi 
- \I \bar{\Psi} \mathcal{D}_{\mu}^{\dag} \gamma^{\mu} \Psi 
\right) 
-m \bar{\Psi} \Psi 
\label{equ:Lag}
\end{equation}
with the Dirac matrices $\gamma^\mu$ and the covariant derivatives
$\mathcal{D}_{\mu} =  \partial_{\mu} +{\rm i} q A_{\mu} $ and 
$\mathcal{D}_{\mu}^{\dag} = \overset{\leftharpoonup} 
{\partial_{\mu}} -{\rm i} q A_{\mu} $. 
As usual in the Furry picture, the Dirac spinors $\bar {\Psi}$ and ${\Psi}$
are treated as dynamical quantum fields whereas the electromagnetic 
vector potential $A_\mu$ is an external field, i.e., a c-number. 
As is well known, by doing so we neglect higher-order interactions, 
e.g., radiative emission, backreaction or electron-electron coupling 
\cite{Fauth, PhysRevD.87.105006,Bloch, PhysRevD.87.125035, PhysRevSTAB.14.054401, PhysRevLett.105.080402}.


Considering bilinear forms of the Dirac spinors $\hat{\bar\Psi}$ and ${\hat\Psi}$
at different space-time points $x_1^\mu$ and $x_2^\mu$, we may transform them to 
center-of-mass ${\mathfrak r}^\mu=(x_1^\mu+x_2^\mu)/2$ and relative coordinates 
${\mathfrak s}^\mu=x_1^\mu-x_2^\mu$ which yields the generalized density operator 
\begin{equation}
\hat{\mathcal C}_{\alpha \beta}\left({\mathfrak r},{\mathfrak s}\right) 
= \mathcal U_A\left({\mathfrak r},{\mathfrak s}\right) 
\left[ 
\hat{\bar\Psi}_\beta \left({\mathfrak r}-{\mathfrak s}/2\right), 
\hat\Psi_\alpha \left({\mathfrak r}+{\mathfrak s}/2\right) 
\right], 
\label{equ:C}
\end{equation}
where the Wilson line factor
\begin{equation}
\mathcal U_A\left({\mathfrak r},{\mathfrak s}\right) 
%
%
= \exp \left( \mathrm{i}q \int_{-1/2}^{1/2} {\rm d} 
\xi \ {\mathfrak s}^\mu A_\mu \left({\mathfrak r}+\xi{\mathfrak s}\right) \right)
\label{equ:U}
\end{equation}
is implemented to ensure gauge-invariance. 

The Fourier transform of this quantity~\eqref{equ:C} with respect to the 
relative coordinate ${\mathfrak s}^\mu=x_1^\mu-x_2^\mu$ yields the 
covariant Wigner operator \cite{Vasak:1987umA} in $2+1$ dimensions
\begin{align}
\hat{\mathcal W}_{\alpha \beta} \left( {\mathfrak r}, p \right) 
= 
\frac{1}{2} \int {\rm d}^3 {\mathfrak s} 
\ \mathrm{e}^{\mathrm{i} p_\mu {\mathfrak s}^\mu} 
\  \hat{\mathcal C}_{\alpha \beta} \left( {\mathfrak r},{\mathfrak s} \right), 
\label{equ:W}
\end{align}
where $p_\mu$ can be identified as the kinetic or mechanical momentum 
because the vector potential $A_\mu$ is already contained in the Wilson line 
factor~\eqref{equ:U}. 
Thus $\hat{\mathcal W}_{\alpha \beta}$ represents a kinetic quantity defined 
in the particles' coordinate-momentum phase-space. 
 
The time evolution of the Wigner operator is determined by the Dirac equation. 
In order to formulate transport equations, we take its expectation value in 
the initial vacuum state $|0\rangle$ which yields the Wigner function 
\begin{equation} 
 \mathbbm{W}
 \left( {\mathfrak r}, p \right) = 
 \langle 0 | \hat{\mathcal W} \left( {\mathfrak r}, p \right) | 0 \rangle,
\end{equation} 
where we omitted the indices for the sake of simplicity. 

We expand the Wigner function in Dirac bilinears in an irreducible representation 
\begin{equation}
 \mathbbm{W}
\left( {\mathfrak r} , p \right) = \frac{1}{2} \left( \mathbbm{1} \mathbbm{S} + \gamma^{\mu} \mathbbm{V}_{\mu} \right),
\end{equation}
where the gamma matrices are (in 2+1 dimensions)
given in terms of the Pauli matrices
\begin{equation}
 \gamma^0 = \sigma_3,\ \gamma^1 = i \sigma_1,\ \gamma^2 = -i \sigma_2.
\end{equation}

\paragraph{Transport equations}

Projection on equal-time 
\begin{equation}
 \mathbbm{w} (t, \boldsymbol{x}, \boldsymbol{p})  
 = 
 \int \frac{{\rm d}p_0}{2 \pi} \, \mathbbm{W} ( {\mathfrak r},p),
\end{equation}
with ${\mathfrak r}=(t,\boldsymbol{x})$ 
gives rise to a closed set of differential equations describing the time 
evolution of particle distributions, namely mass $ \mathbbm{s}$, 
charge $ \mathbbm{v}_\mathbb{0}$ and current 
$ \boldsymbol{\mathbbm {v}}$ density \cite{Vasak:1987umA,BB,Ochs}. 
For a potential of the form~\eqref{collision},
within QED$_{2+1}$ and one set of $2$-spinors we obtain on the basis of $
 p_x = k_x,~
 p_y = k_y + q \int {\rm d} \xi \ A_y \left( x + \xi {\mathfrak s}_x ,t \right) $
\cite{Kohlf, PhysRevD.101.096009, KOHLFURST2016371} 
\begin{alignat}{8}
    & \partial_t \ \mathbbm{s} && && && +2 k_x \ \mathbbm{v}_2 && -2 \Pi \ \mathbbm{v}_1 &&= 0, \label{eq_5_1} \\     
    & \partial_t \ \mathbbm{v}_\mathbb{0} &&+ D \ \mathbbm{v}_2 && + \partial_x \ \mathbbm{v}_1 && && &&= 0,  \\ 
    & \partial_t \ \mathbbm{v}_1 && && + \partial_x \ \mathbbm{v}_0 && && +2 \Pi \ \mathbbm{s} &&= +2m \ \mathbbm{v}_2, \\
    & \partial_t \ \mathbbm{v}_2 && + D \ \mathbbm{v}_0 && && -2 k_x \ \mathbbm{s} && &&= -2m \ \mathbbm{v}_1, \label{eq_5_2}      
\end{alignat}
with pseudo-differential operators \cite{KohlfurstNew}    
\begin{alignat}{6}
 & D && =  & +\I q \ \mathcal{F}^{-1}_{k_x} \ \left[ A_y \left( x+ \frac{{\mathfrak s}_x}{2},t \right) - A_y \left( x- \frac{{\mathfrak s}_x}{2},t \right) \right] \mathcal{F}_{k_x}, && \label{eq_diff1} \\ 
 & \Pi && = k_y &+ \frac{q}{2} \ \mathcal{F}^{-1}_{k_x} \ \left[ A_y \left( x+ \frac{{\mathfrak s}_x}{2},t \right) + A_y \left( x- \frac{{\mathfrak s}_x}{2},t \right) \right] \mathcal{F}_{k_x}. \label{eq_diff2} &&   
\end{alignat}  
Here, the Fourier operators $\mathcal{F}_{k_x}$ transform from canonical momentum space $k_x$ to relative coordinate space ${\mathfrak s}_x$. 
As a matter of fact, Eqs. \eqref{eq_5_1}-\eqref{eq_5_2} describe a partial differential equation in $t,x$ and $k_x$ with the (conserved) canonical momentum $k_y$ serving as an external parameter. 
 
In order to determine the pair production rate from an initial vacuum state, we employ initial conditions of the form
\begin{alignat}{7}
& \mathbbm{s}_{\rm vac} \left(\boldsymbol{k} \right) = -\frac{m}{\sqrt{m^2 +
   \boldsymbol{k}^2}}, \qquad \label{eq:IC} && 
\boldsymbol{\mathbbm{v}}_{\rm vac} \left(\boldsymbol{k} \right) = -\frac{
   \boldsymbol{k}}{\sqrt{m^2 + \boldsymbol{k}^2}}, \qquad
& \mathbbm{v}_\mathbb{0} {}_{\rm vac} = 0  . 
\end{alignat}
Evaluating the transport equations on the basis of these initial conditions, 
we obtain the particle distribution function at asymptotic times 
($t \to \pm \infty$) 
\begin{equation}
n \left( \boldsymbol{x}, \boldsymbol{k} \right) 
= 
\frac{m \left( \mathbbm{s}-\mathbbm{s}_{\rm vac} \right) + \boldsymbol{k} \cdot \left(
\boldsymbol{\mathbbm{v}}-\boldsymbol{\mathbbm{v}}_{\rm vac} \right)}{2\sqrt{m^2+\boldsymbol{k}^2}}.
 \label{equ:n2}
\end{equation}

The total particle number is given by
\begin{equation}
 N = \int \ \frac{{\rm d}^2k \ {\rm d}^2x}{(2 \pi)^2} \  n \left( \boldsymbol{x}, \boldsymbol{k} \right).
\end{equation}

\paragraph{Spatially homogeneous approximation}

In the spatially homogeneous field approximation we calculate a fully time-dependent, but spatially localized pair production rate ($x=0$) on the basis of Eqs. \eqref{eq_5_1}-\eqref{eq_5_2}, see also \cite{footnote2}. 

At $x=0$ the vector potential takes on the form $\f{A}(t)=2f(ct)\f{e}_y$, 
see Eq. \eqref{Gaussian}. Consequently, the differential operators \eqref{eq_diff1}-\eqref{eq_diff2} simplify to ordinary factors, $D=0$ and $\Pi=q A_y$. Furthermore, within an entirely localized description of particle creation, particle propagation can be neglected thus derivatives with respect to spatial coordinates vanish.

As a result, the corresponding equations of motion take on the much simpler form
\begin{alignat}{8}
& \partial_t \ {\mathbbm{s}}^{\rm SHA} && && && +2 k_x \ {\mathbbm{v}}_2^{\rm SHA} && -2 \left(k_y + qA_y(t) \right) \ {\mathbbm{v}}_1^{\rm SHA} &&= 0, \label{eq_3_1} \\     
& \partial_t \ {\mathbbm{v}}_1^{\rm SHA} && && && && +2 \left(k_y + qA_y(t) \right) \ {\mathbbm{s}}^{\rm SHA} &&= +2m \ {\mathbbm{v}}_2^{\rm SHA}, \\
& \partial_t \ {\mathbbm{v}}_2^{\rm SHA} && && && -2 k_x \ {\mathbbm{s}}^{\rm SHA} && &&= -2m \ {\mathbbm{v}}_1^{\rm SHA}, \label{eq_3_2}  
\end{alignat}
where we used the superscript ${\rm SHA}$ (spatially homogeneous approximation) to indicate that these components do not depend on $x$. Initial conditions remain unchanged, see Eq \eqref{eq:IC}. Equations \eqref{eq_3_1}-\eqref{eq_3_2} can also be derived employing a quantum kinetic approach \cite{Hebenstreit, Kluger, Schmidt}.

\paragraph{Locally constant field approximation}

Within the locally constant field approximation we assume instantaneous particle creation. While this approximation fails to capture time-memory effects, e.g., photon absorptive processes, it is expected to provide correct particle numbers if spatial and temporal variations in the employed fields are sufficiently smooth.

For a field of the form $\f{A}(t,\f{r})$ we model the source term heuristically after the Sauter-Schwinger effect in constant fields \cite{Nikishov2} in $2+1$ dimensions \cite{Lin}
\begin{equation}
 {\mathcal S} = \frac{\lvert q \rvert^{3/2} \ a(t,x)^{3/2}}{(2 \pi)^2} \exp \left( -\frac{\pi m^2}{\lvert q \rvert a(t,x)} \right), \label{Eq:LCFA}
\end{equation}
with $a(t,x) = \sqrt{ \lvert {\mathcal F(t,x)} \rvert - {\mathcal F(t,x)} }$
and ${\mathcal F}(t,x) = -1/2 \ \big( E(t,x)^2 - B(t,x)^2 \big)$.
In order to obtain the total particle number integrations over all coordinates 
are in order, $N_{\rm LCFA} = \int {\rm d}t \ \int {\rm d}x \ {\mathcal S}(t,x)$.


\section{C. Curvature contribution}

\begin{figure}[t]
 \begin{center}
  \includegraphics[trim={1cm 0cm 3cm 1cm},clip,height=0.15\textheight]{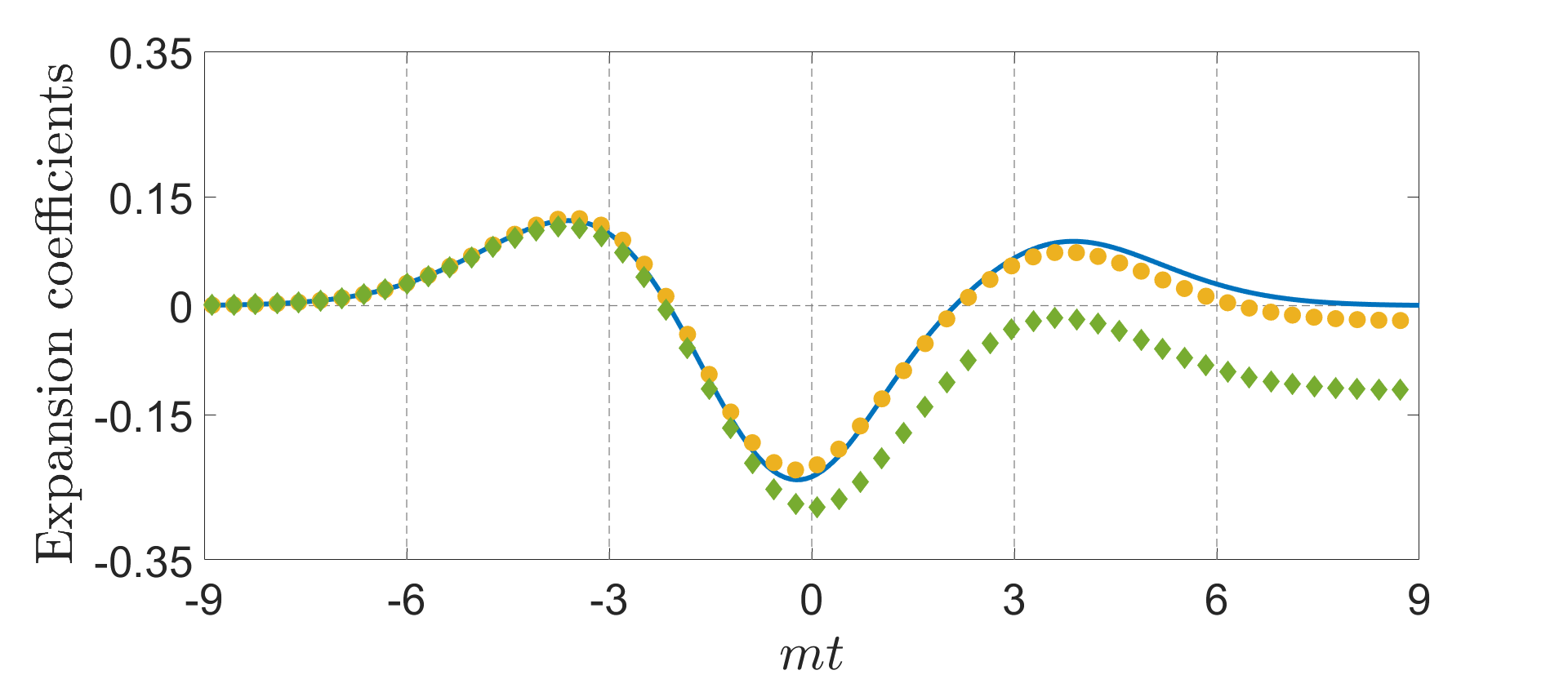} 
  \includegraphics[trim={1cm 0cm 3cm 1cm},clip,height=0.15\textheight]{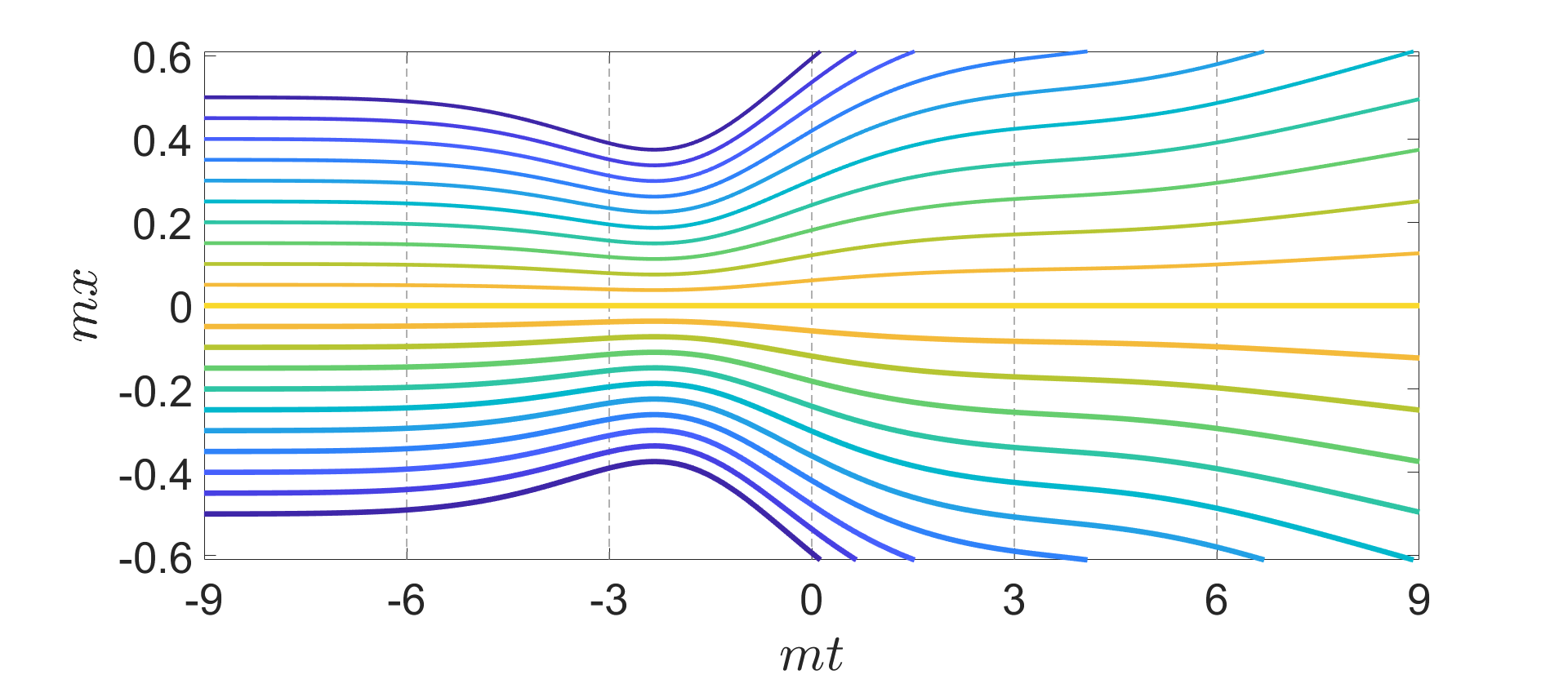}   
  \includegraphics[trim={1cm 0cm 3cm 1cm},clip,height=0.15\textheight]{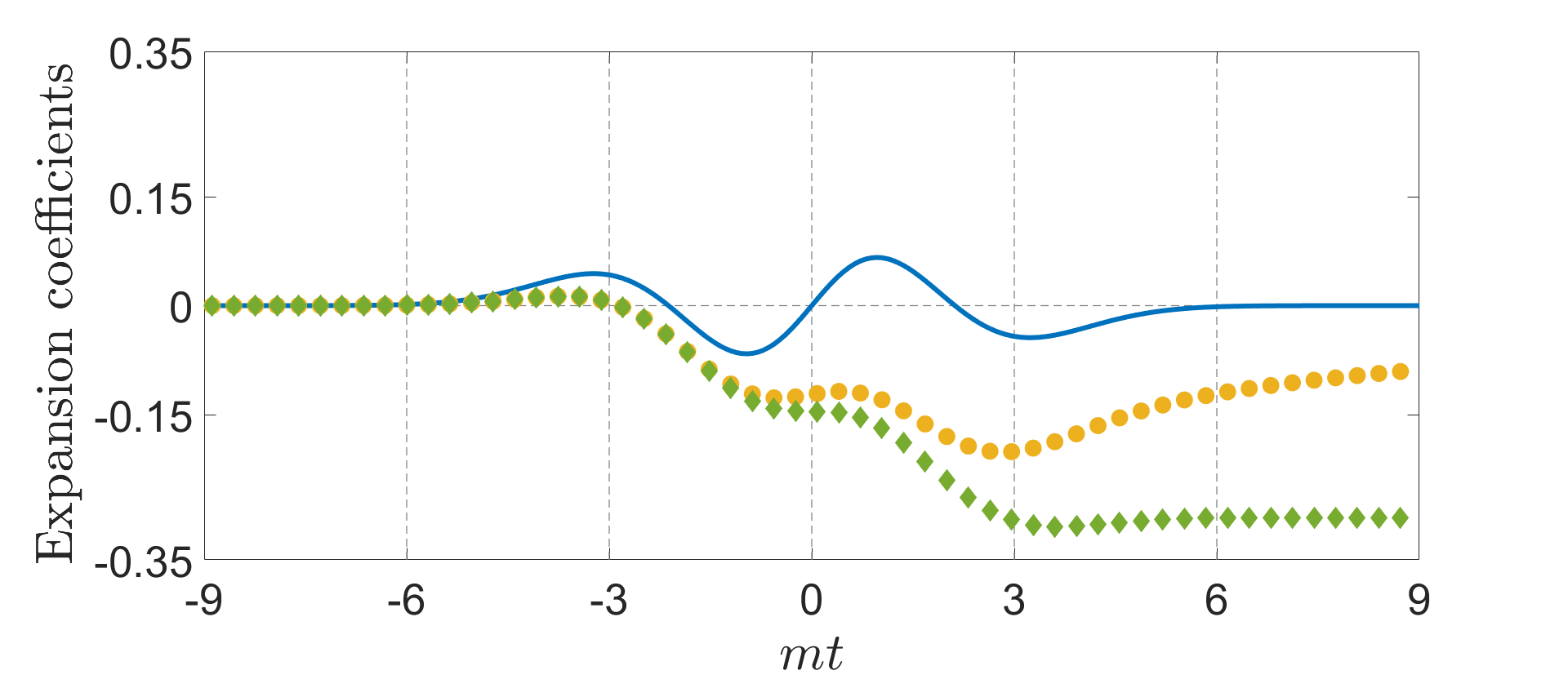} 
  \includegraphics[trim={1cm 0cm 3cm 1cm},clip,height=0.15\textheight]{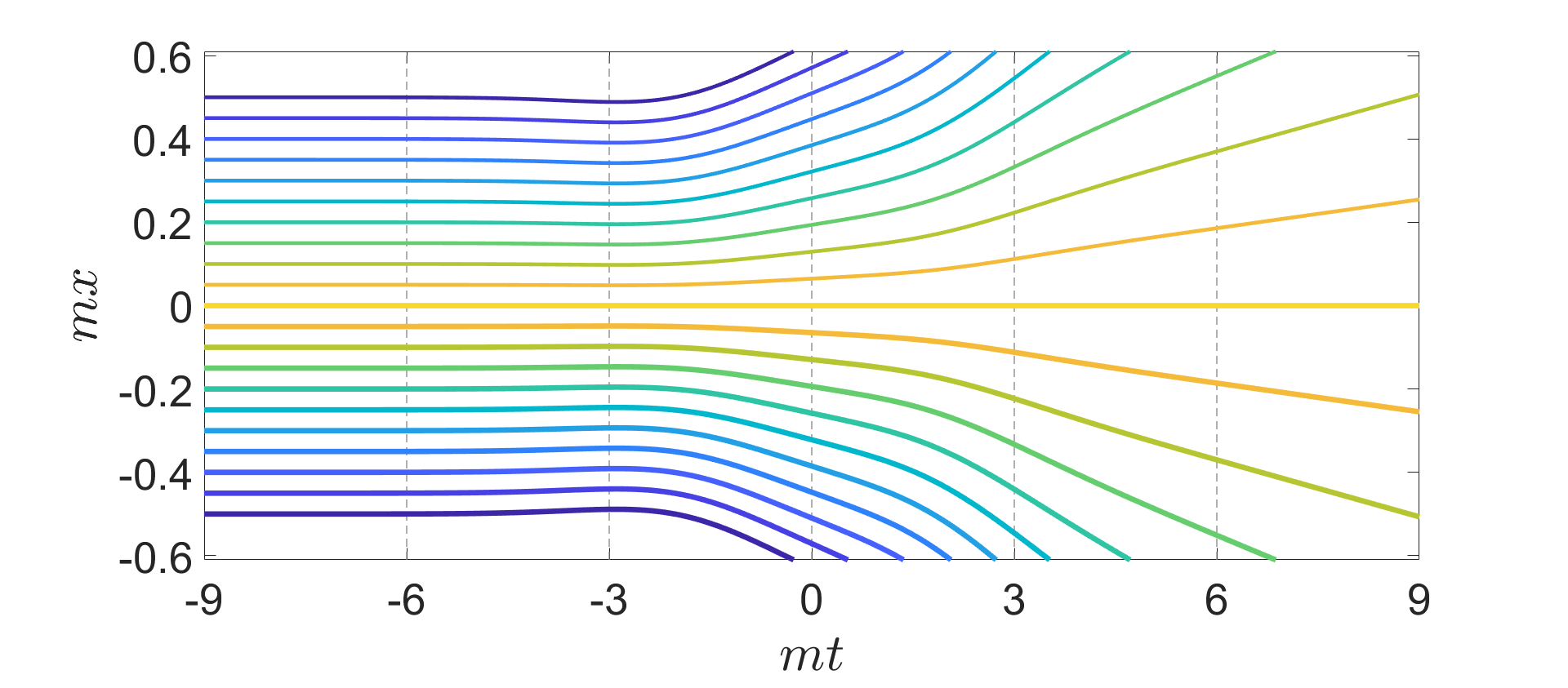}   
  \includegraphics[trim={1cm 0cm 3cm 1cm},clip,height=0.15\textheight]{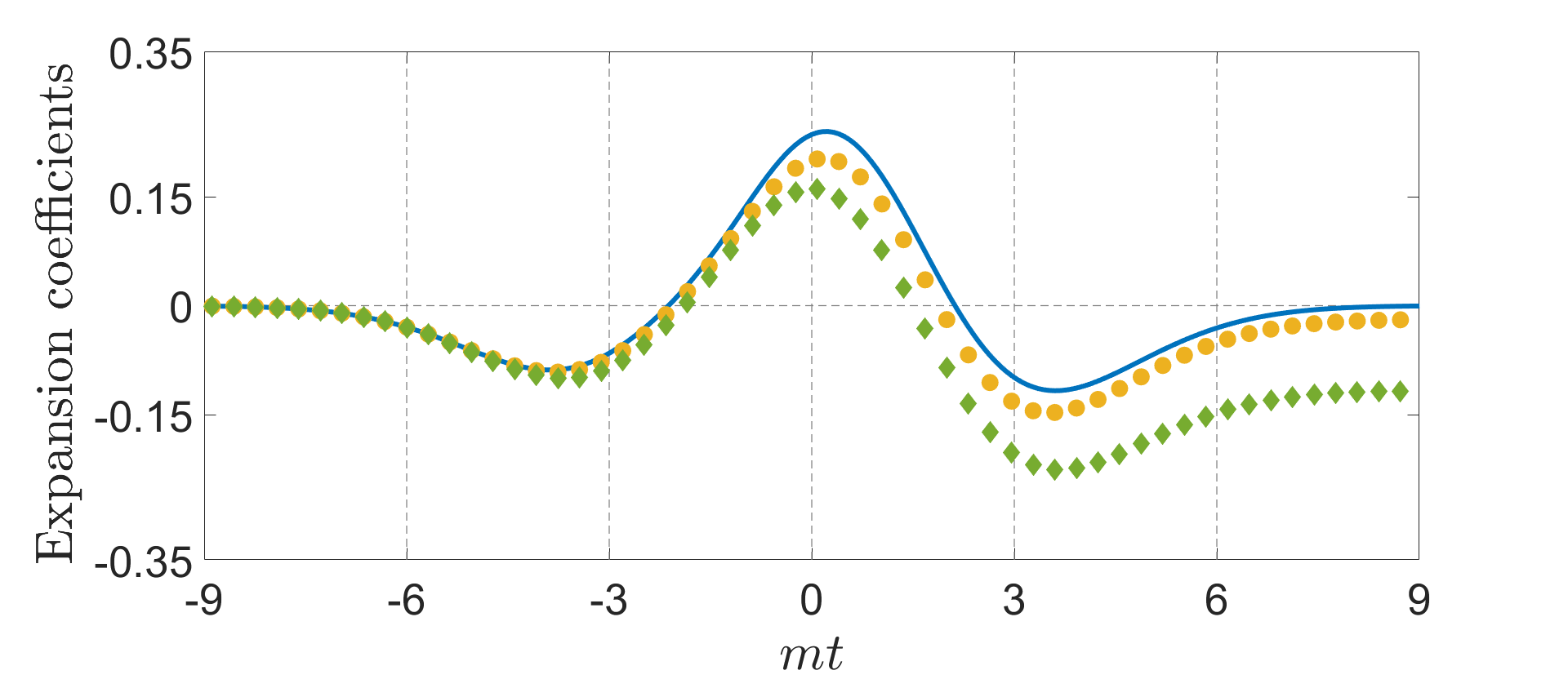} 
  \includegraphics[trim={1cm 0cm 3cm 1cm},clip,height=0.15\textheight]{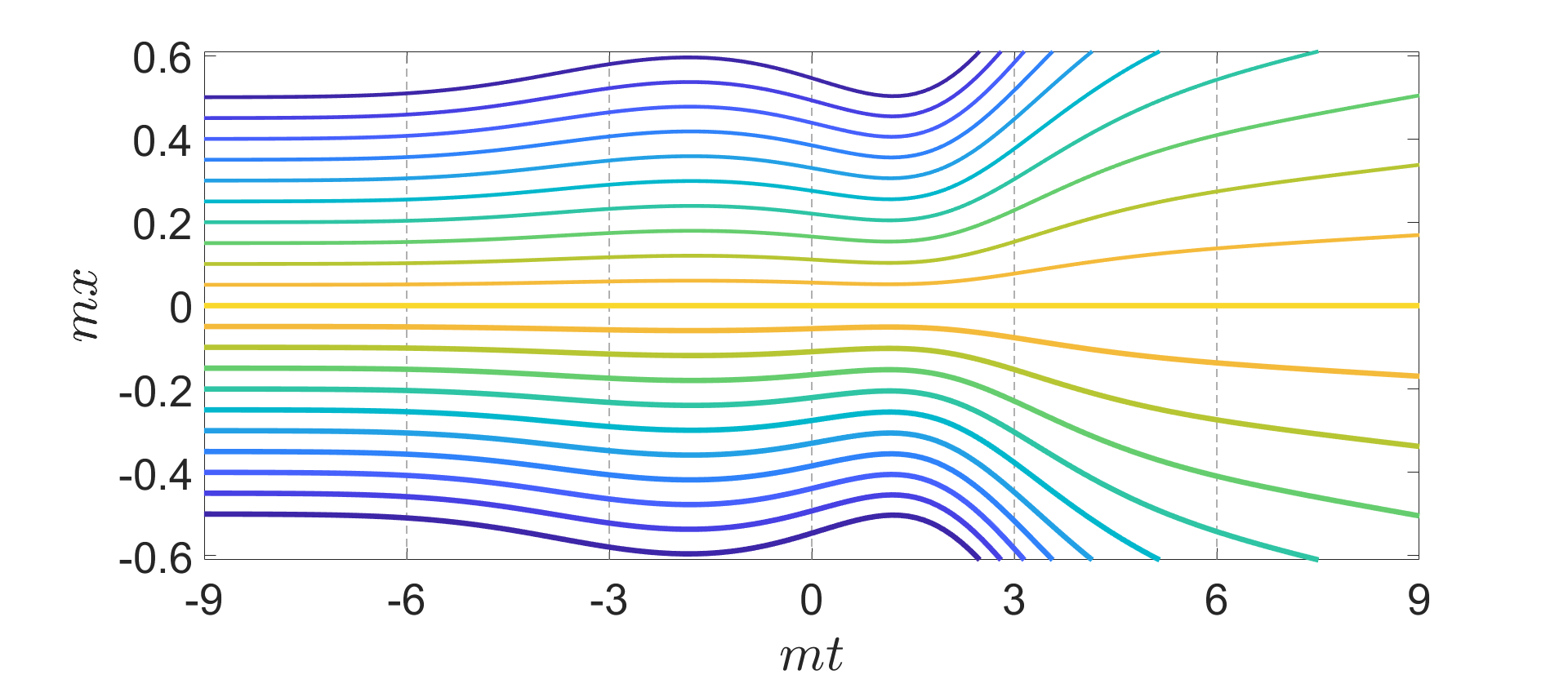}   
 \end{center}
 \caption{Left panel: Time evolution of $\ddot s_0$ (blue line) in  
 Eq.~\eqref{s_0} and the curvature $s_2$ in Eq.~\eqref{curvature} 
 in the linear 
 (green diamonds) and non-linear (orange circles) form, respectively. 
 Data are obtained for the profile~\eqref{Gaussian} with $E=E_S/3$, 
 $\omega=m/3$ and $k_y=-m$ (top), $k_y=0$ (middle), $k_y=m$ (bottom).
 In the right panel, we display a set of characteristic curves \cite{Oertel}  
 for the same parameters in order to visualize the focussing/de-focussing
 effects -- where one may observe a strong correlation to the behavior of $s_2$,
 as expected from the discussion after Eq.~(10).
 }
 \label{fig_Evo_s}
\end{figure} 

\begin{figure}[t]
\includegraphics[width=0.7\textwidth]{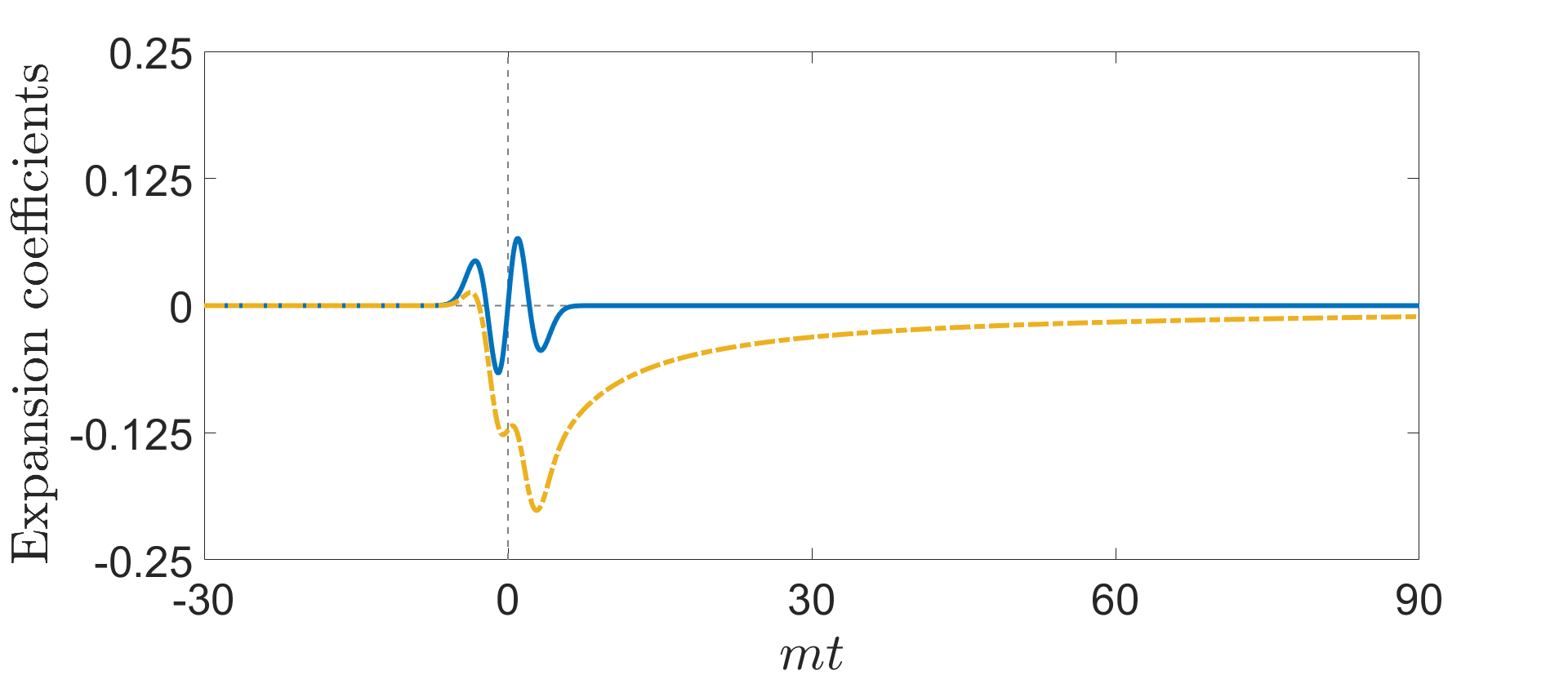} 
\includegraphics[width=0.7\textwidth]{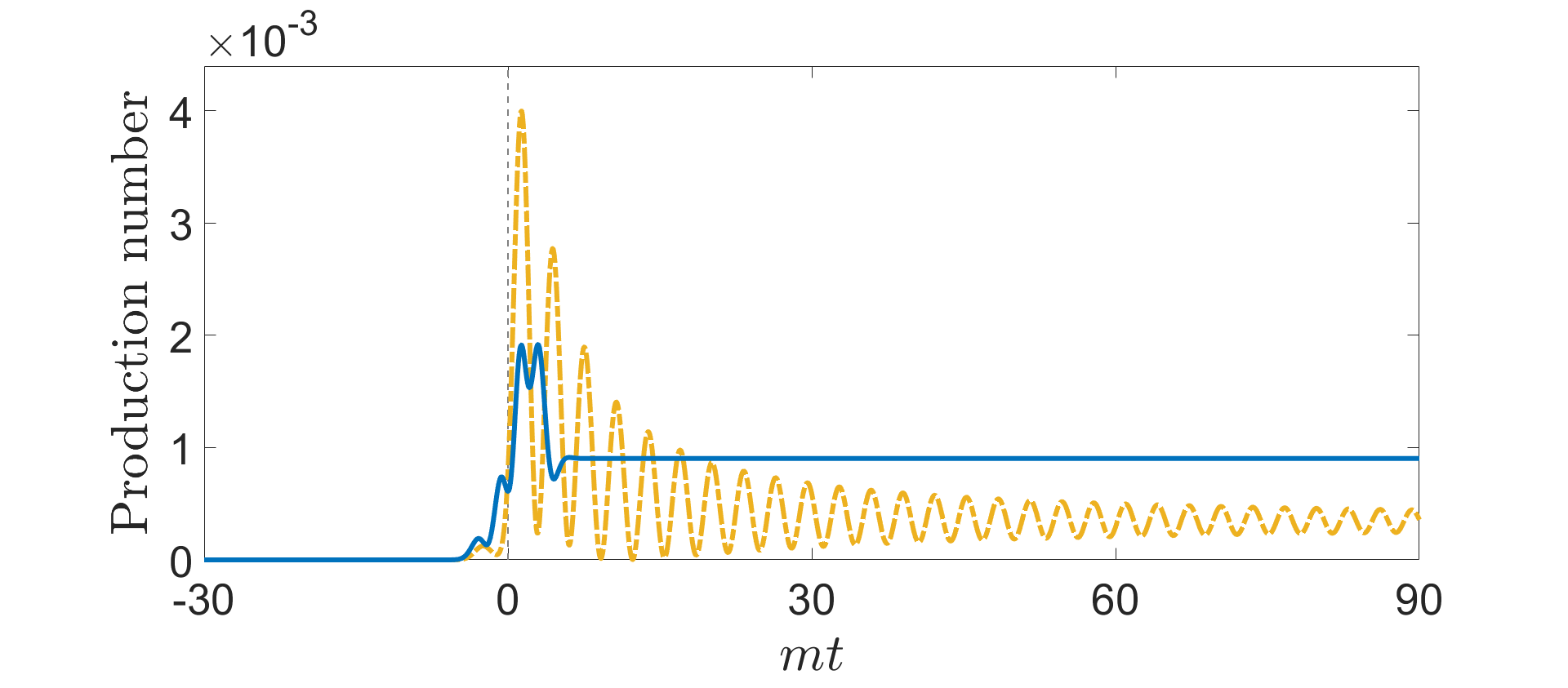} 
\caption{Time evolution of $\ddot s_0$ (solid blue line) 
in Eq.~\eqref{s_0} and the curvature $s_2$ (dashed orange line)
in Eq.~\eqref{curvature} as in Fig.~\ref{fig_Evo_s} (top); 
the quantity $\lvert R \rvert ^2$ associated with the rate of particle creation, 
obtained in the spatially homogeneous field approximation (solid blue line), 
i.e., just from $\ddot s_0$, 
in comparison to the full solution (dashed orange line)
obtained by replacing $\ddot s_0\to\Box s=\ddot s_0-s_2$ taking into account
the focusing/de-focusing corrections (bottom). 
Data are obtained for the profile~\eqref{Gaussian} with $E=E_S/3$, 
$\omega=m/3$ and $k_y=0$.}
\label{caustics}
\end{figure}

Solving Eq.~\eqref{curvature} numerically, we plot the evolution of the 
curvature $s_2$ (starting at zero) in comparison to $\ddot s_0$ for $E=E_S/3$, 
$\omega=m/3$ and the values $k_y=0$ and $k_y=\pm m$ in Fig.~\ref{fig_Evo_s}. 

As our first observation, we find no blow-up singularities for these parameters.
%
%
Furthermore, we compare the full solution of the non-linear differential 
equation~\eqref{curvature} with its linearized approximation obtained by 
neglecting the nonlinearity $s_2^2$ in Eq.~\eqref{curvature}. 
We find reasonably good agreement in the 
time interval relevant for pair creation.
%
%
Note that the potential blow-up singularities are caused by the nonlinearity 
$s_2^2$ in Eq.~\eqref{curvature} and thus never occur
in the linearized solution.

For $k_y=\pm m$, we see that the curvature $s_2(t)$ lies very close to 
$\ddot s_0(t)$. 
As a result, this curvature contribution (stemming from the magnetic field)
almost completely cancels the term $\ddot s_0(t)$ in the pre-factor $\Box s$ 
of the evolution equations~\eqref{evolution-Bogoliubov} for the Bogoliubov
coefficients, leading to a significantly reduced pair creation. 
This cancellation does not occur for $k_y=0$,  
but the pair creation probability is still reduced by roughly a factor of two,
see Fig.~\ref{caustics} and Section~G below.


Whether the differences between the full Dirac-Heisenberg-Wigner data 
and the results of the spatially homogeneous field approximation observed 
in Fig.~\ref{spectrum} below, most notably the ``shoulders'' of the curve 
with $\omega=m/3$ starting around $|k_y|\approx m/2$, can be traced back 
(at least partially) to this cancellation for large $|k_y|$ should be 
investigated in future studies.

For the sake of completeness, we also examine the characteristic curves for a 
better visualization of focusing and de-focusing effects created by the 
inhomogeneity in the field. 
To this end, we employ the method of characteristics in order to turn 
Eq.~\eqref{square-root} into a family of first-order, ordinary differential 
equations
\begin{align}
 \dot x_0(\tau) = \frac{2}{m} p_0(\tau), \,
 \dot x_1(\tau) = -\frac{2}{m} p_1(\tau), \,
 \dot p_0(\tau) = \frac{2}{m} \left(k_y +q A_y\right) \left(q E_y\right), \,
 \dot p_1(\tau) = \frac{2}{m} \left(k_y +q A_y\right) \left(q B_z\right),  
\end{align}
with $p_0 = \partial_t s$ and $p_1 = \partial_x s$, respectively. 
The parameter $\tau$ specifies the location on a distinctive characteristic 
curve determined by a particular initial condition. 
A set of solutions obtained by varying the initial starting position 
$x_{1,{\rm init}}$ is displayed in Fig.~\ref{fig_Evo_s}.
It is apparent that there is a strong correlation between $s_2$ being 
positive/negative and the curves coming closer or being dispersed, respectively. 


\section{D. Particle spectra}

\begin{figure}[b]
\includegraphics[trim={3cm 0 3cm 0},clip,width=0.49\textwidth]
{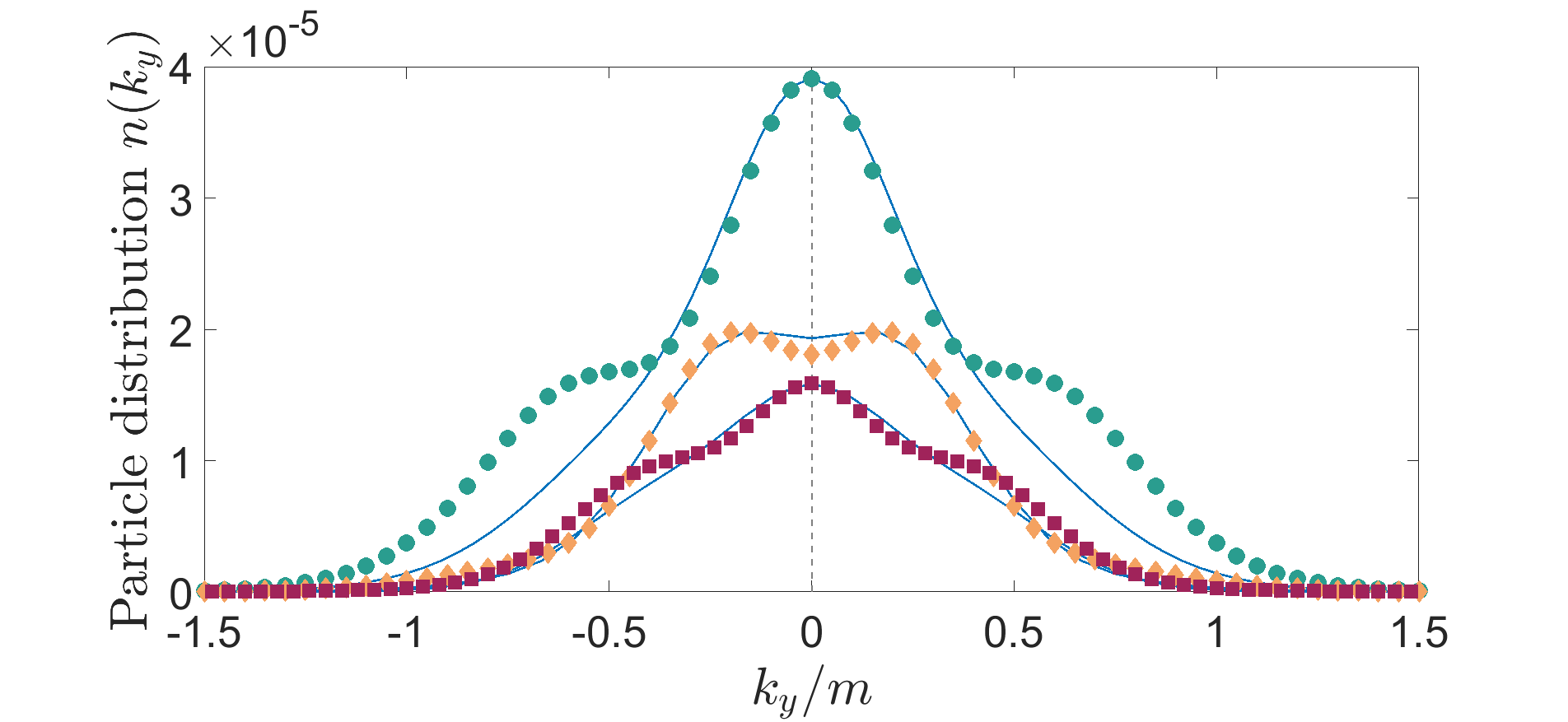}
\caption{Plot of the pair-creation spectra obtained by the spatially 
homogeneous field approximation for the profile~\eqref{Gaussian} with 
$E=E_S/3$ and 
$\omega=m/3$ (green circles), 
$\omega=m/4$ (orange diamonds), and 
$\omega=m/5$ (red squares). 
The data have been rescaled by a constant factor in order to compensate 
the overestimation mentioned above. 
For comparison, the blue curves correspond to the results obtained through 
the Dirac-Heisenberg-Wigner approach.}
\label{spectrum}
\end{figure}

While the world-line instanton technique -- at least in its simplest form 
discussed above -- only yields the total pair creation 
probability~\eqref{instanton}, the WKB and Dirac-Heisenberg-Wigner approaches 
allow direct access to the particle spectra. 
Since $k_x$ is not a conserved quantity in our case~\eqref{collision},
we focus on the $k_y$-dependence. 
Figure~\ref{spectrum} displays exemplary spectra for the frequency values
$\omega=m/3$, $\omega=m/4$ and $\omega=m/5$, where we have used the same 
scenario as in Fig.~\ref{vergleich}, i.e., the profile~\eqref{Gaussian} 
with $E=E_S/3$. 
Shown are the results of the Dirac-Heisenberg-Wigner approach in comparison
to the spatially homogeneous field approximation (after re-scaling), which 
are basically the spectra produced by a purely time-dependent field. 

We find that the spectra obtained by the two approaches match quite well,
especially for small $k_y$ (consistent with the considerations in the 
previous Section), but there are also distinctive differences. 
Most notably, the spatially homogeneous field approximation predicts two 
``shoulders'' starting around $k_y\approx\pm m/2$ in the spectrum for 
$\omega=m/3$, which are not reproduced by the Dirac-Heisenberg-Wigner 
approach.
These differences might be explainable by the additional curvature 
contributions $s_2$ discussed in the previous Section, but further 
investigations are necessary to settle this point. 


As another interesting feature, both approaches agree on a plateau 
or even dip at small $k_y$ in the spectrum for $\omega=m/4$. 
This is quite remarkable since it seems to contradict the standard 
expectation that the spectrum should have its maximum at $k_y=0$.
Assuming that the particles are predominantly created at $t=0$
(i.e., the maximum of the electric field) and with vanishing initial 
velocity $v_y(t=0)=0$, one would indeed expect the spectrum to have 
its maximum at $k_y=0$ in view of $A_y(t=0)=A_y(t\to\pm\infty)=0$. 
Investigating the reasons for this apparent contribution 
(e.g., interference phenomena, see also 
\cite{PhysRevLett.108.030401, PhysRevD.98.056009, PhysRevD.83.065028}) 
should be the subject of further studies. 

\begin{sidewaysfigure}[t]
 \begin{center}   
 \includegraphics[trim={4cm 0 2.5cm 0},clip,width=0.99\textwidth]{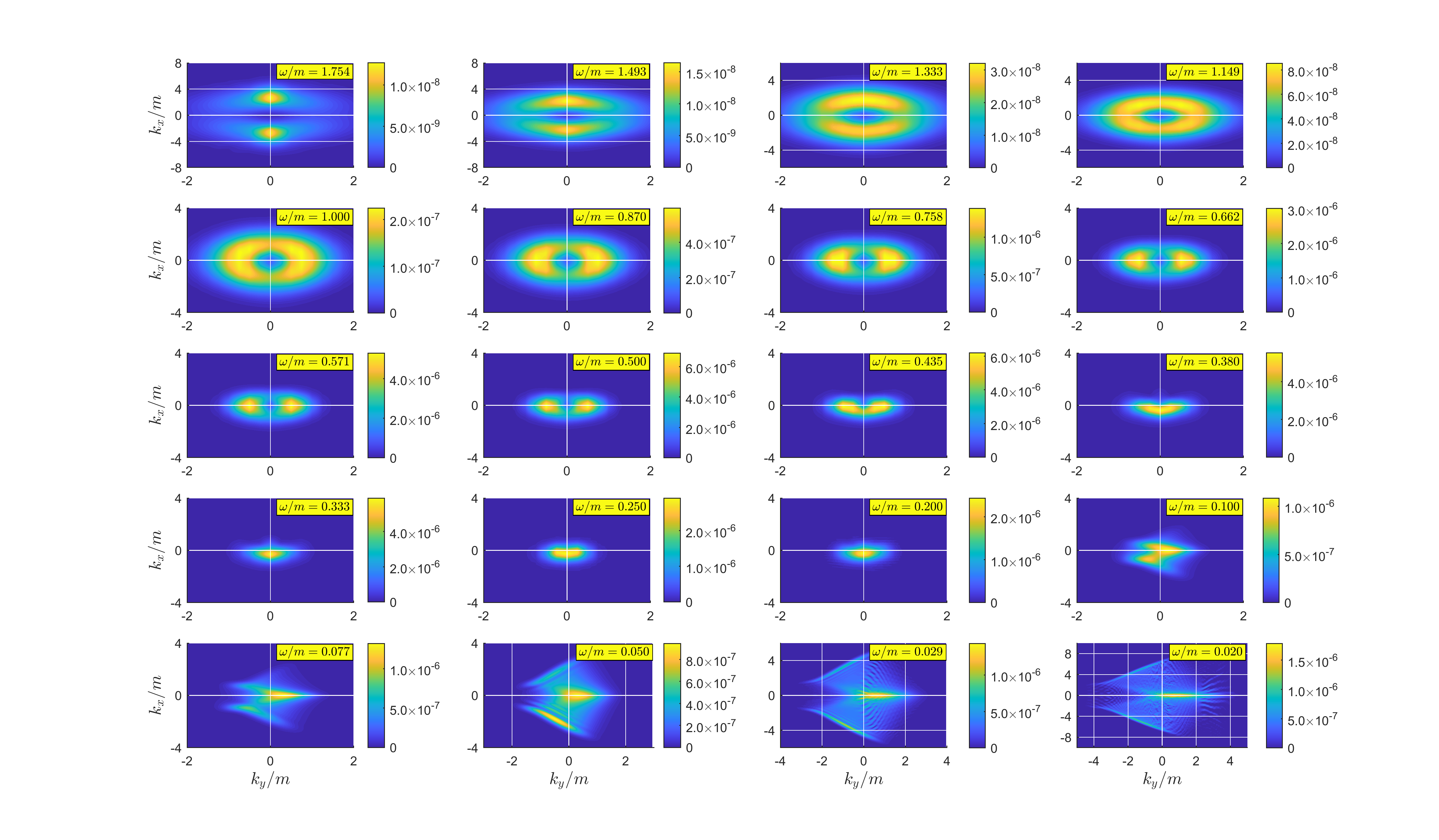}   
 \end{center}
 \vspace{-1cm}
 \caption{Particle momentum spectra obtained from solving the  
 Dirac-Heisenberg-Wigner transport equations for a field profile of two identical, 
 colliding laser pulses \eqref{collision}. 
 The field strength is given by $E=E_S/3$, the different pulse frequencies are 
 displayed as insets. 
 Through integration over the momentum space variables $k_x$ and $k_y$ 
 the mean number of created particles is obtained, cf.~the orange  circles in 
 Fig.~\ref{vergleich}.}
 \label{fig:DHW_fpxpz}   
\end{sidewaysfigure}

For the sake of completeness, we illustrate the two-dimensional spectra 
(displaying the dependence on $k_x$ and $k_y$) 
obtained through solving the transport equations \eqref{eq_5_1}-\eqref{eq_5_2} 
for a variety of parameters in 
Fig.~\ref{fig:DHW_fpxpz}. 
Note that integrating these spectra over $k_x$ yields the blue curves in 
Fig.~\ref{spectrum} while another integration over $k_y$ then gives the 
total particle number, i.e., the orange curve in Fig.~\ref{vergleich}.


\section{E. Threshold effects}

\begin{figure}[h]
 \begin{center}   
  \includegraphics[trim={1.5cm 0cm 3cm 0.75cm},clip,width=0.49\textwidth]{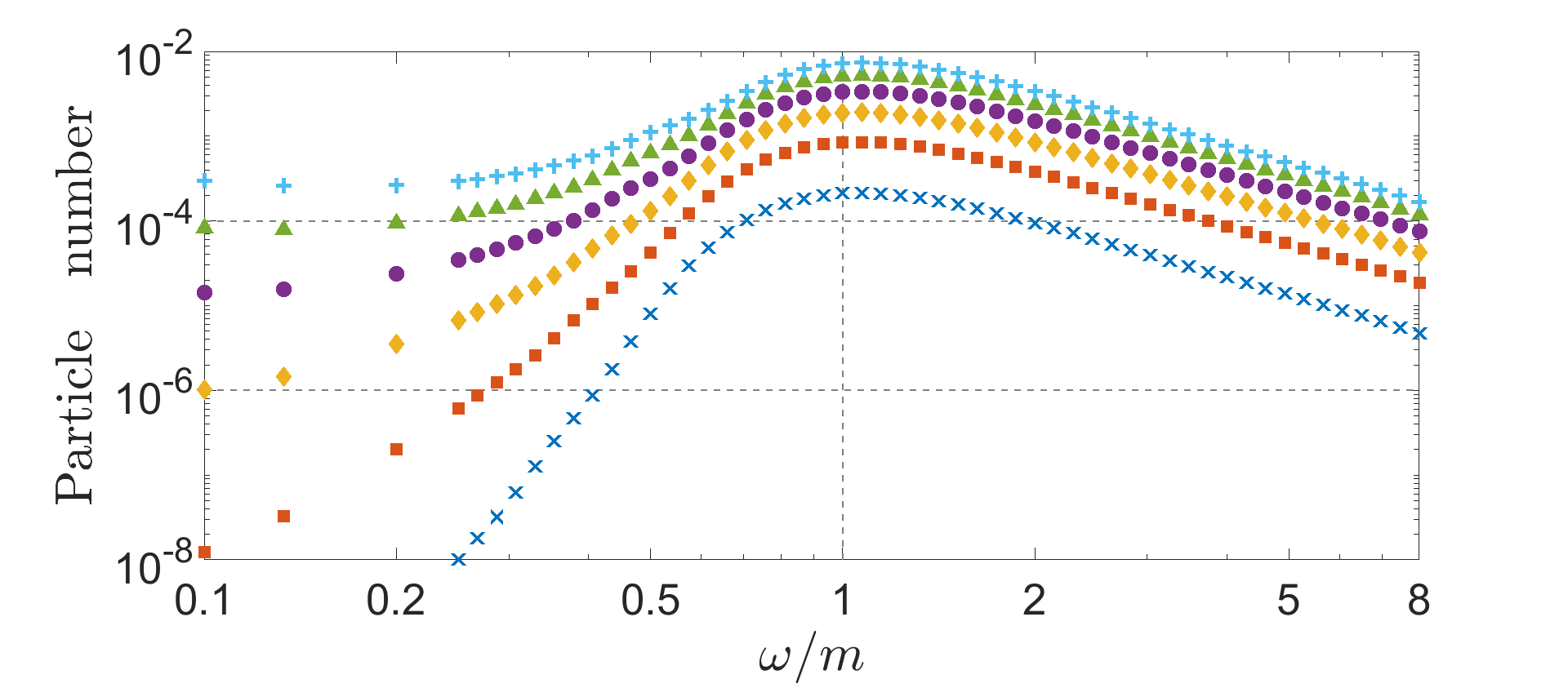} 	
  \includegraphics[trim={1.5cm 0cm 3cm 0.75cm},clip,width=0.49\textwidth]{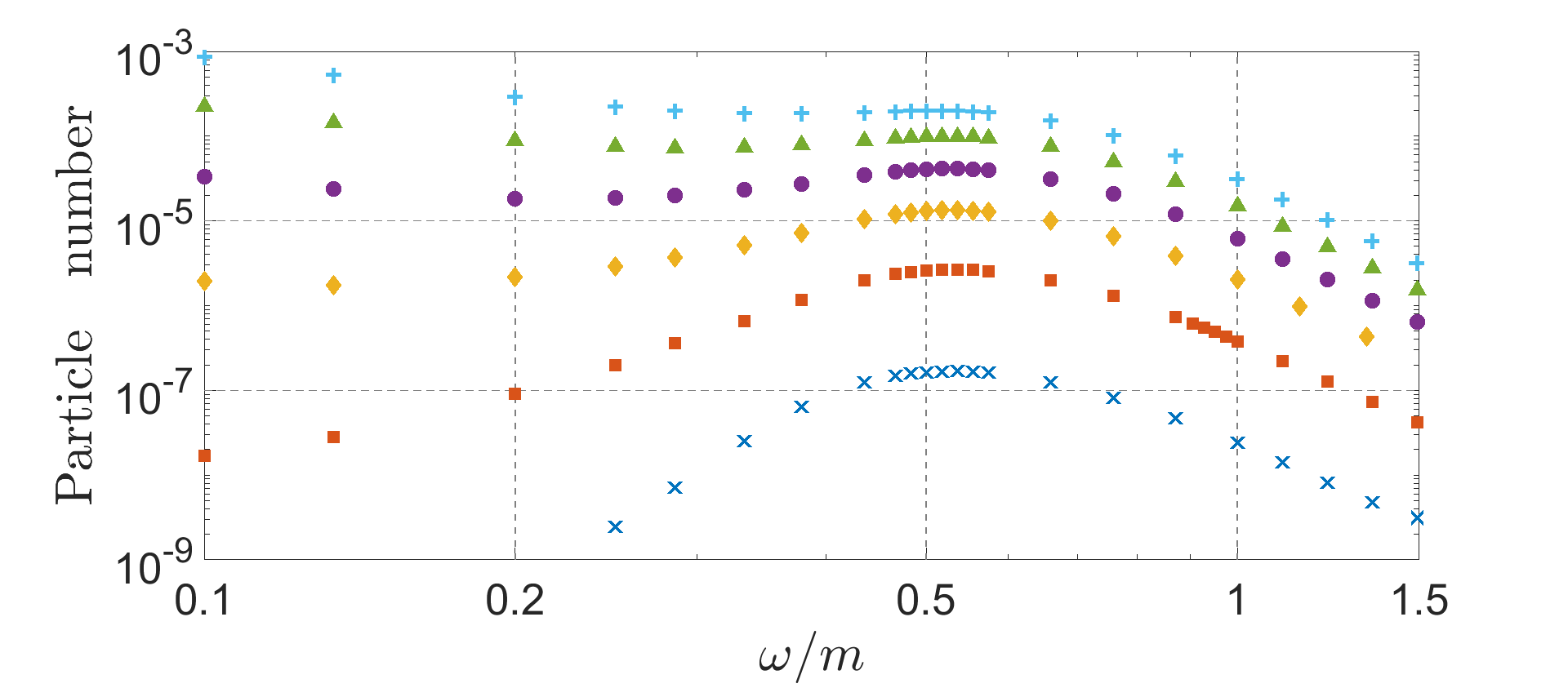} 	 	 
 \end{center}
 \caption{Mean number of created particles as a function of $\omega$ for various 
 field strengths $E/E_S\in\{0.1,0.2,\ldots, 0.6\}$, for a purely time-dependent 
 profile \eqref{Gaussian} at the left and the collision scenario \eqref{collision}
 at the right. 
 }
 \label{figScaling}   
\end{figure}

As a working hypothesis, we interpret the peaks observed in Fig.~\ref{vergleich}
as threshold effects marking the transition from the non-perturbative to the 
perturbative regime. 
In order to test this hypothesis, we study the dependence of these peaks on the 
maximum electric field strength $E$ in Fig.~\ref{figScaling}.

To lowest (non-vanishing) order perturbation theory, one would expect an 
$E^4$-scaling of the mean particle number in the collision 
scenario~\eqref{collision} where two photons collide to form an 
electron-positron pair.
Such an $E^4$-scaling is indeed consistent with the dependence of the peak 
heights as well as the behavior of the curves for larger frequencies in 
Fig.~\ref{figScaling} (right). 
For smaller frequencies, however, the curves display deviations from this 
scaling, indicating the failure of lowest-order perturbation theory. 

For purely time-dependent fields, one would expect an $E^2$-scaling instead,
as already the first-order amplitude in $E$ can account for pair creation. 
Again, this scaling is consistent with the peak heights as well as the 
behavior for larger frequencies in Fig.~\ref{figScaling} (left), 
while the dependence at smaller frequencies deviates. 

These perturbative arguments can also help to understand the different 
locations of the peaks.
While the collision of two photons with frequency $\Omega \geq m$ may create $e^+e^-$-pairs in the collision scenario, a purely time-dependent field 
must contain frequency components with $\Omega \geq 2m$ in order to obtain a non-zero first-order amplitude in $E$.
Consistently, the peak of the collision scenario occurs at lower 
frequencies than that for purely time-dependent fields. 

Note that making these points more precise is complicated by the fact that 
the Fourier transform of the field profile~\eqref{Gaussian} yields a rather 
broad frequency spectrum which has its maximum at a frequency of $\Omega = 2\omega$. 
Elucidating this issue should be the subject of further studies. 

\section{F. Blow-up singularities}

Let us study the presence or absence of blow-up singularities in the curvature 
contribution $s_2(t)$ in more detail. 
First, we focus on the case $k_y=k_z=0$ for simplicity and consider large 
Keldysh parameters $\gamma\gg1$, i.e., $qE\ll m\omega$. 
In this limit, we have $qA_y\ll m$ and using 
$\partial_x^2A_y=\partial_t^2A_y=\ord(\omega E)$, 
we find that $s_2$ scales with $s_2=\ord(qE/\gamma)$, i.e., $s_2\ll qE$. 
Altogether, we may simplify Eq.~(10) via 
\bea 
\dot s_2
=
\left.
\frac{s_2^2+\left[k_y+qA_y\right]q\partial_x^2A_y
}{\sqrt{m^2+\left[k_y+qA_y\right]^2+k_z^2}} 
\right\rvert_{x=0}
\approx 
\left.\frac{q^2A_y\ddot A_y}{m}\right\rvert_{x=0}
=
-\left.\frac{q^2(\dot A_y)^2}{m}\right\rvert_{x=0}
+ \left.
\frac{d}{dt}\left(\frac{q^2A_y\dot A_y}{m}\right) \right\rvert_{x=0}
\,. 
\ea
Apart from the total derivative in the last term, 
we find that $\dot s_2$ is negative. 
Thus, $s_2(t)$ may oscillate during the collision of the two pulses, 
but assumes a negative value afterwards. 
The subsequent evolution is then governed by the non-linearity 
$\dot s_2=s_2^2/m$ which implies that $s_2(t)$ is slowly approaching the 
$t$-axis from below as $1/t$. 
Blow-up singularities would occur if $s_2$ was positive after the collision 
and could be visualized as self-focusing, but they are absent in this case 
and the de-focusing effects dominate. 

Actually, as we have observed in Fig.~\ref{fig_Evo_s}, the linearization 
(i.e., neglect of $s_2^2$) provides a fairly good approximation even for 
moderate values of $k_y=\ord(m)$ and for Keldysh parameters $\gamma$ 
of order unity. 
Thus, if we rewrite $\dot s_2$ as 
\bea 
\dot s_2
=
-
\left.
\frac{q^2\dot A_y^2m^2}{\left(m^2+\left[k_y+qA_y\right]^2+k_z^2\right)^{3/2}}
\right\rvert_{x=0}
+ \left.
\frac{d}{dt}\left(
\frac{q\dot A_y\left[k_y+qA_y\right]}{\sqrt{m^2+\left[k_y+qA_y\right]^2+k_z^2}}
\right) \right\rvert_{x=0}
+\ord(s_2^2)
\,, 
\ea
we again find a negative value of $s_2$ after the collision, provided that the 
non-linearity $s_2^2$ can be neglected during the collision. 
Thus, for moderate values of $\gamma$ and $k_y$ 
(note that $k_z$ can simply be absorbed into $m$), 
we basically get the same picture as above, i.e., 
an approximately linear evolution of $s_2(t)$ during the collision 
resulting in a negative value of $s_2$ after the collision, 
which then implies a slow $1/t$-decay of $|s_2|$ governed by the 
non-linear evolution. 

Consistent with these analytic approximations, we only found blow-up 
singularities in our numerical simulation for very large $k_y$ and/or 
for extremely long pulses (which can be treated via the locally constant 
field approximation).

\section{G. Bogoliubov coefficients}

The Bogoliubov coefficients $\alpha(t,x=0)$ and $\beta(t,x=0)$ along the 
symmetry plane can either be obtained by solving Eqs.~(15) directly or from 
the Riccati equation $\dot R=\Box s(e^{+2is}-R^2e^{-2is})/(2\chi)$ together with 
their normalization. 
For our choice of the eigenvectors $\f{u}_\pm=(1,\pm i\chi)^T$, 
the normalization of the Bogoliubov coefficients along the symmetry plane can 
be derived from Eqs.~(15) and reads 
\bea
\label{normalization}
\left|\alpha(t,x=0)\right|^2-\left|\beta(t,x=0)\right|^2
=
\exp\left\{
-\frac12\int\limits_{-\infty}^t dt'\,
\left.\frac{\Box s}{\chi}\right\rvert_{t',x=0}
\right\}
\,.
\ea
For purely time-dependent fields $\Box s=\ddot s$, the integrand 
$\ddot s/\chi=\ddot s/\dot s$ in Eq.~\eqref{normalization}
is a total derivative and thus we 
recover the well-known $1/\sqrt{\chi}$ normalization. 
Including the spatial curvature $s_2$ also incorporates 
focusing/de-focusing effects.  
Actually, inserting the dependence $s_2(t)\propto1/t$ for late times 
(as explained above), the right-hand side of Eq.~\eqref{normalization}
behaves as $1/\sqrt{t}$ which corresponds to the spread of the wave packets. 
Obviously, this does not imply that the number of created particles decreases 
-- their number should be constant after the collision is over -- 
it just means that they do not stay at $x=0$ but eventually move away. 
In order to factor out this trivial spreading effect, we consider the ratio 
$R=\beta/\alpha$ where the overall normalization~\eqref{normalization}
cancels. 
%

More specifically, starting with a normalized wave-packet of purely positive frequency (corresponding to the $\alpha$ coefficient), the probability of particle creation is determined by the norm of the final negative-frequency 
part of the wave-packet (corresponding to the $\beta$ coefficient). 
Obviously, the trivial spreading of the wave-packet does not change this 
norm.
Thus, in order to determine the pair-creation probability, we introduce 
the normalized Bogoliubov coefficients $\tilde\alpha=\alpha/\cal N$ 
and $\tilde\beta=\beta/\cal N$ where $\cal N$ is given by 
Eq.~\eqref{normalization} via $|\alpha|^2-|\beta|^2={\cal N}^2$. 
These normalized Bogoliubov coefficients satisfy the usual unitarity 
relation $|\tilde\alpha|^2-|\tilde\beta|^2=1$ and have the same ratio 
$R=\tilde\beta/\tilde\alpha=\beta/\alpha$.
The pair-creation probability is then given by $|\tilde\beta|^2$ which 
can be obtained from $R$ via 
\bea
|\tilde\beta|^2=\frac{|R|^2}{1-|R|^2}
\,.
\ea
For small $|R|^2\ll1$ as in Fig.~\ref{caustics} 
(implying $|\tilde\alpha|^2\gg|\tilde\beta|^2$) 
this simplifies to $|\tilde\beta|^2\approx|R|^2$. 


%
%
%
%
%

\section{H. Numerical Simulation}

The transport equations \eqref{eq_5_1}-\eqref{eq_5_2} have been solved 
according to the blueprint presented in \cite{Kohlf}. 
Thus, the time evolution is performed on the basis of a high-order 
Dormand-Prince Runge-Kutta integrator with adaptive time-stepping. 
This includes an artificial super-exponential adiabatic turn on/off 
of the field in order to avoid high-frequency components spoiling the 
simulation corresponding to a computational initial time $t_i \approx 40/m$ 
even for profiles where $\omega \sim \ord({m})$. 
Derivative operators \eqref{eq_diff1}-\eqref{eq_diff2} are evaluated 
employing discrete (inverse) Fourier transforms \cite{Frigo}.

\onecolumngrid
\twocolumngrid 



\begin{thebibliography}{99}
\bibitem{footnote2}
 Note that the spatially homogeneous field approximation applied here is not 
 synonymous with the locally homogeneous approximation used in, for example, 
 \cite{HebenstreitSelf}. 
 %
 In the latter, the field $A(t,x)$ at each space point $x$ is mapped onto a 
 purely time-dependent field and the resulting pair creation density is then 
 integrated over all positions $x$ (in analogy to the locally constant field 
 approximation, just with $x$ instead of $x$ and $t$). 
 %
 Thus, this approximation scheme completely neglects the impact of the 
 magnetic field component \cite{Kohlf}. 
 %
 As a result, it would yield a finite particle creation rate even for a 
 single propagating pulse  $\f{A}(t,\f{r})=A_y(t,x)\f{e}_y=f(t-x)\f{e}_y$
 which is not correct.
\bibitem{electron-bunch}
 On a different length scale, one could also consider tightly focused electron bunches instead of the nuclei, see also \cite{PhysRevLett.122.190404}.
\end{thebibliography}
\begin{thebibliography}{99}
\bibitem{footnote-prefactor} 
 The investigation of the pre-factor within the world-line 
 instanton method for the colliding-pulse profile~\eqref{collision} will be 
 the subject of a forthcoming publication.
\end{thebibliography}
\end{document}